\documentclass[
superscriptaddress,
 reprint,
 amsmath,amssymb,
 aps,%onecolumn
]{revtex4-2}

\def\final{0}

\usepackage{verbatim}
\usepackage{rotating, booktabs}  % For table-rotating 
\usepackage{braket}
\usepackage{graphics}
\usepackage{graphicx}
\usepackage{hyperref}
	\hypersetup{hypertex= true, colorlinks = true,linkcolor = blue, anchorcolor = blue, citecolor= blue}
\usepackage{color, xcolor}
\usepackage{upgreek}  % greek font
\usepackage{dcolumn}
\newcolumntype{z}[1]{D{.}{.}{#1}}
\usepackage{subfigure}
\usepackage{float}    % Floats,Figures and Captions
\usepackage{etoolbox}
 \usepackage{threeparttable}
 \usepackage{multirow}
 \usepackage{float}
 \usepackage{makecell}
 \usepackage{booktabs, tabularx, colortbl}
 \usepackage{array}
 \usepackage{enumerate}

\usepackage[linesnumbered, ruled]{algorithm2e}
\SetKwInput{KwOutput}{Output}              % set the Output
\SetKwInOut{Input}{Input}
\SetKwInput{Output}{Output} 
\usepackage{comment}

\usepackage{xargs}
\usepackage[colorinlistoftodos,prependcaption]{todonotes}
\newtheorem{definitionenv}{Definition}
\newtheorem{lemmaenv}[definitionenv]{Lemma}
\newtheorem{theoremenv}[definitionenv]{Theorem}
\newtheorem{corollaryenv}[definitionenv]{Corollary}
\newtheorem{propositionenv}[definitionenv]{Proposition}
\newtheorem{conjectureenv}[definitionenv]{Conjecture}
\newtheorem{remarkenv}[definitionenv]{Remark}
\newenvironment{remark}{\begin{remarkenv}\rm}{\end{remarkenv}}
\newcommand{\br}{\begin{remark}}
	\newcommand{\er}{\end{remark}}

\newtheorem{exampleenv}{Example}
\newtheorem{app-lemmaenv}[section]{Lemma}

\newenvironment{definition}{\begin{definitionenv}\rm}{\end{definitionenv}}
\newenvironment{lemma}{\begin{lemmaenv}\rm}{\end{lemmaenv}}
\newenvironment{theorem}{\begin{theoremenv}\rm}{\end{theoremenv}}
\newenvironment{corollary}{\begin{corollaryenv}\rm}{\end{corollaryenv}}
\newenvironment{example}{\begin{exampleenv}\rm}{\end{exampleenv}}
\newenvironment{proposition}{\begin{propositionenv}\rm}{\end{propositionenv}}
\newenvironment{conjecture}{\begin{conjectureenv}\rm}{\end{conjectureenv}}
\newenvironment{app-lemma}{\begin{app-lemmaenv}\rm}{\end{app-lemmaenv}}

\newcommand{\bd}{\begin{definition}}
	\newcommand{\ed}{\end{definition}}
\newcommand{\bl}{\begin{lemma}}
	\newcommand{\el}{\end{lemma}}
\newcommand{\elp}{\hspace*{\fill} $\Box$
\end{lemma}}
\newcommand{\bt}{\begin{theorem}}
\newcommand{\et}{\end{theorem}}
\newcommand{\etp}{\hspace*{\fill} $\Box$
\end{theorem}}
\newcommand{\bc}{\begin{corollary}}
\newcommand{\ec}{\end{corollary}}
\newcommand{\ecp}{\hspace*{\fill} $\Box$
\end{corollary}}
\newcommand{\bcj}{\begin{conjecture}}
\newcommand{\ecj}{\end{conjecture}}

\newcommand{\be}{\begin{example}}
\newcommand{\ee}{\end{example}}
\newcommand{\eep}{\hspace*{\fill} $\Box$
\end{example}}
\newcommand{\bp}{\begin{proposition}}
\newcommand{\ep}{\end{proposition}}
\newcommand{\epp}{%\hspace*{\fill} $\Box$
\end{proposition}}

\newcommand{\cS}{\mathcal{S}}

\newcommand{\codepar}[1]{\ensuremath{[\![#1]\!]}}

\ifnum\final=0
\newcommand{\mynote}[2]{{\color{#1} \marginpar{\tiny #2}}}
\newcommand{\mybignote}[2]{{\color{#1} $\langle \langle$ #2$\rangle \rangle$}}
\newcommandx{\rednote}[2][1=]{\todo[linecolor=red,backgroundcolor=red!25,bordercolor=red,#1]{#2}}
\newcommandx{\bluenote}[2][1=]{\todo[linecolor=blue,backgroundcolor=blue!25,bordercolor=blue,#1]{#2}}
\newcommandx{\yellownote}[2][1=]{\todo[linecolor=yellow,backgroundcolor=yellow!25,bordercolor=yellow,#1]{#2}}
\newcommandx{\greennote}[2][1=]{\todo[inline,linecolor=olive,backgroundcolor=green!25,bordercolor=olive,#1]{#2}}

\else
\newcommand{\mynote}[2]{}
\newcommand{\mybignote}[2]{}
\newcommand{\rednote}[2][1=]{}
\newcommand{\bluenote}[2][1=]{}
\newcommand{\greennote}[2][1=]{}
\newcommand{\yellownote}[2][1=]{}

\fi

\begin{document}

 \title{Reducing Quantum Error Correction Overhead with Versatile Flag-Sharing  Syndrome Extraction Circuits}

\author{Pei-Hao Liou}
 \email{phliou.ee08@nycu.edu.tw}
 \affiliation{Institute of Communications Engineering, National Yang Ming Chiao Tung University,  Hsinchu 30010, Taiwan} 
\author{Ching-Yi Lai}%
 \email{cylai@nycu.edu.tw}
 \affiliation{Institute of Communications Engineering, National Yang Ming Chiao Tung University,  Hsinchu 30010, Taiwan}
 \affiliation{Physics Division, National Center for Theoretical Sciences, Taipei 10617, Taiwan}

\date{\today}% It is always \today, today,

\begin{abstract}

Given that quantum error correction processes are unreliable, an efficient error syndrome extraction circuit should use fewer ancillary qubits, quantum gates, and measurements, while maintaining low circuit depth,
to minimizing the circuit area, roughly defined as the product of circuit depth and the number of physical qubits. 
 We propose to design parallel flagged syndrome extraction with shared flag qubits for   quantum stabilizer codes.  Versatile parallelization techniques are employed to minimize the required circuit area, thereby improving the error threshold and overall performance.
Specifically,   all the measurement outcomes in multiple rounds of syndrome extraction 
are integrated into a lookup table decoder,  allowing us to parallelize  multiple stabilizer measurements with shared flag qubits.
We present flag-sharing and fully parallel schemes for  the \codepar{17,1,5} and \codepar{19,1,5} Calderbank-Shor-Steane (CSS) codes. This methodology extends to the \codepar{5,1,3}  non-CSS code, achieving the minimum known circuit area. 
Numerical simulations have demonstrated improved pseudothresholds for these codes by up to an order of magnitude compared to previous schemes in the literature.

% 	\keywords{flag qubit \and Parallel syndrome extraction \and More}
\end{abstract}

\maketitle

\section{Introduction}
 
Quantum computers have the potential to solve certain problems much more efficiently than classical computers by leveraging the properties of quantum entanglement and parallel computation. However, implementing large-scale quantum algorithms requires a significant number of qubits, the fundamental units of quantum computation. Qubits are highly susceptible to decoherence due to external environmental interference, which poses a significant challenge to reliable quantum computation~\cite{Preskill18, AL18,  CBV17, takita17}. To achieve large-scale and reliable quantum computing, it is essential to find methods to protect qubits from the effects of decoherence or to develop more stable qubits. One promising approach to address this challenge is through fault-tolerant quantum computation (FTQC)~\cite{AB97,Shor96,DS96,KLZ96,Got97,Pre98c,Steane97,Ste99N,ND05}.

FTQC uses quantum error correcting codes (QEC) \cite{Shor95, Steane97} to encode physical qubits into logical qubits, ensuring that errors can be corrected when they occur. When the error rate of physical gates is below a certain threshold, arbitrarily precise quantum computation can be achieved \cite{AB97,AGP06,AGP08}. However,
to perform active quantum error correction, a fault-tolerant (FT) syndrome extraction (SE) procedure is implemented, which requires a significant number of ancilla qubits, two-qubit quantum gates, and single-qubit measurements.  Additionally, because these quantum components are unreliable, repeated SE procedures are necessary to obtain reliable recovery information, adding another layer of complexity to the process. The considerable resource demands of FTQC present a significant challenge for its implementation.

Many SE approaches have been presented, including the Shor, Steane, and Knill procedures \cite{Shor96,Steane97,Knill05}, which require the preparation of specific ancillary states. For the Steane and Knill SE procedures, FT state preparations are typically performed in advance~\cite{LZB17,ZLB+20}. For the Shor-style SE, dealing with unreliable measurements often involves repeated measurements or using redundant syndrome measurements~\cite{ALB20,KCL21}. 
Another approach is flagged SE~\cite{CR18a,CR18b,CB18}, where a minimal number of ancilla qubits are delicately designed to detect multiple error propagation events. This method utilizes adaptive lookup-table decoding to achieve fault-tolerant quantum error correction.
Recently, it has been shown that an adaptive SE procedure can reduce the required number of repeated measurements on average by utilizing information across consecutive rounds for error correction~\cite{TPB23}. 
Building on this approach, Pato et al. \cite{PTH+24} further enhanced its efficiency by developing a suite of optimization tools capable of reducing space and time overhead in SE. These tools can also be integrated with the flagged SE method to improve overall performance.

It is crucial to minimize the resources required in flagged FTQC. Reichardt designed an SE circuit for the \codepar{7,1,3} code that utilizes measurement qubits as flag qubits, enabling simultaneous measurement of multiple stabilizers and reducing circuit depth~\cite{Rei20}. Liou and Lai introduced parallel SE circuits with shared flag qubits, which measure multiple stabilizers of the same type  for Calderbank-Shor-Steane (CSS) type stabilizer codes of distance three~\cite{LL23}. Following this, Du et al. \cite{DML+24} developed parallel schemes with shared flag qubits for \codepar{17,1,5} and \codepar{19,1,5} codes based on the decoding algorithm proposed in~\cite{CB18}.

 In this paper, we propose a methodology for deriving efficient SE circuits for general CSS codes. We define the notion of the effective circuit area of an SE circuit as the total number of two-qubit gates, ancilla preparations, measurements, and idle qubits in the circuit, scaled by their relative error rates. An SE circuit with a low effective circuit area will be more efficient in FTQC, leading to a better error threshold.
 
 Achieving good parallelism in SE to minimize the number of idle qubits can effectively reduce the effective circuit area. Traditional flag schemes utilize serial SE circuits to measure all the stabilizer generators~\cite{CR18b,CB18,BSE+23}. We show that these multiple SE stages in the serial scheme can be parallelized by interleaving the CNOTs in their order on the data qubits, thus enabling depth reduction. Therefore, a serial flagged SE scheme for stabilizers of the same type implies the existence of a corresponding fully parallel flagged SE scheme. By carefully adjusting the order of the two-qubit gates, it is possible to obtain an SE scheme with minimal circuit depth.

 Using the technique of flag-sharing can reduce the number of ancilla qubits and measurements at the cost of additional two-qubit gates and potentially more idle qubits. We provide design criteria for determining whether some stabilizer SE circuits can be merged using shared flags. Moreover, we show that a sequential flag-sharing SE scheme also implies the existence of a corresponding fully parallel version. When the measurement error rate dominates, it is preferable to reduce the number of measurements by using fewer flag qubits in SE.
 
   \begin{figure}[htbp]
 	\centering
 	\includegraphics[width=0.85\linewidth]{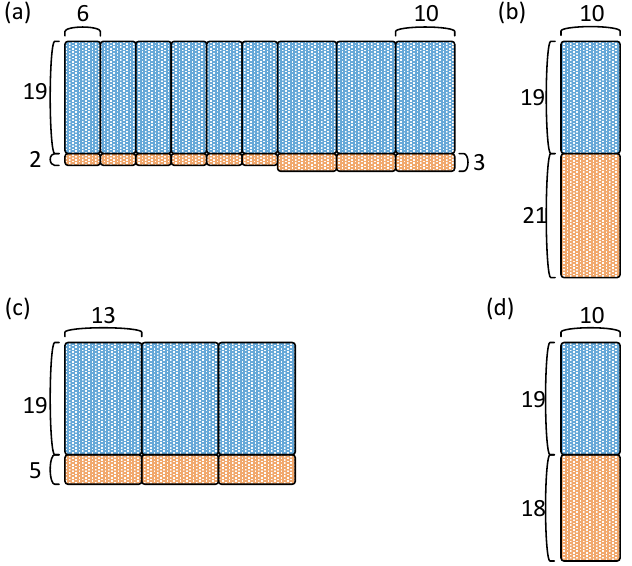}
 	\caption{
 		Versatile SE schemes for the \codepar{19,1,5} code are illustrated:
 		(a) a serial $\mathbf{1}^T$ scheme;
 		(b) a fully parallel $\mathbf{1}$ scheme;
 		(c) a  sequential [3~3~3] flag-sharing scheme;
 		(d) a  sequential [2;~2;~2;~1;~1;~1] flag-sharing scheme.
 		The symbol~$\mathbf{1}$ is understood as an all-ones column vector of appropriate dimension.
 	}\label{fig.tile_1915}
 \end{figure}

Overall, we can derive versatile SE schemes for a given CSS code. We provide examples for the \codepar{17,1,5} and \codepar{19,1,5} codes, featuring low effective circuit areas. For instance, Fig.\ref{fig.tile_1915} illustrates four SE schemes for the \codepar{19,1,5} code, where each scheme is defined by a matrix. Since the \codepar{19,1,5} code has stabilizers of weight 6, a flagged SE using two flag qubits has a depth of 10. Both the fully parallel scheme in Fig.\ref{fig.tile_1915}(b) and the flag-sharing scheme in Fig.~\ref{fig.tile_1915}(d) have been optimized to have a circuit depth of 10 for measuring the entire set of $Z$ ($X$) stabilizers.
Using our schemes, the \codepar{17,1,5} and \codepar{19,1,5} codes exhibit pseudothresholds of $6 \times 10^{-5}$ and $9 \times 10^{-5}$, respectively, when all  components have the same error rates. These results outperform known schemes in the literature~\cite{CB18,DML+24}.

Extending the above-mentioned results to non-CSS codes is more challenging since an SE stage of a stabilizer involves both $X$ and $Z$ components, leading to complex error propagation events. However, we propose versatile sequential flag-sharing SE schemes for the \codepar{5,1,3} code, utilizing its cyclic structure~\cite{LMPZ96}.
In particular, the pseudothresholds of the flag-sharing and fully parallel schemes for the \codepar{5,1,3} code are about $3\times 10^{-4}$, assuming all components have the same error rate. This makes the \codepar{5,1,3} code a competitive candidate for near-term experiments compared to the \codepar{7,1,3} or \codepar{9,1,3} codes, as the \codepar{5,1,3} code requires fewer qubits in total.

In flagged FT QEC, error recovery is performed based on the error syndromes obtained from multiple rounds of SE. Typically, $O(d)$ rounds of flagged or raw SE are sufficient for FT SE for a quantum code of distance $d$. However, by carefully examining the error correction decision tree, it is possible to use fewer than $d$ rounds of SE. We propose using all measurement outcomes as generalized error syndromes to expand the syndrome space and define a unified lookup table. This approach helps reduce the depth of SE schemes and simplifies lookup-table decoding. Importantly, this technique allows us to find the circuits of lowest depth for the aforementioned codes.

Computer simulations on the various schemes are conducted under different error rate scenarios. Our results show that both our flag-sharing and fully parallel schemes outperform known schemes in the literature, as the circuit areas have been optimized in our designs. The resources required for various schemes and their effective circuit areas are summarized in Table~\ref{table.list_area}. Our simulations suggest that the performance of an SE scheme is closely related to its effective circuit area. While one might expect a fully parallel scheme to perform better due to minimized circuit depth and no additional two-qubit gates between different stabilizer measurements, our flag-sharing schemes demonstrate competitive or superior performances due to their lower effective circuit areas.

Recent experiments in quantum computing with atom arrays have focused on achieving parallel execution of entangling gates across multiple qubits. Evered et al. \cite{Evered2023} demonstrated a neutral-atom-based quantum computer capable of high-fidelity entangling operations on multiple qubit pairs simultaneously. Bluvstein et al. \cite{Bluvstein2024} developed a logical quantum processor using reconfigurable atom arrays, enabling concurrent manipulation of multiple atoms with parallel control laser beams.
By harnessing parallel processing capabilities, our parallel SE technique is well-suited for near-term quantum technologies, complementing these recent advancements in quantum computing.

This paper is organized as follows. In Sec.\ref{sec.preliminaries}, we introduce the notion of FTQC and flagged SE. Moving on to Sec.\ref{sec:SE_CSS}, we propose versatile SE schemes for CSS codes, using the \codepar{17,1,5} code as an illustration. In Sec.\ref{sec:SE_nonCSS}, we propose versatile SE schemes for the \codepar{5,1,3} non-CSS code. The QEC procedures for codes with distances 3 or 5 are discussed in Sec.\ref{sec:QEC}. Simulation results on the various SE schemes and their analysis are presented in Sec.\ref{sec.simulation}. Finally, we conclude in Sec.\ref{sec.conclusion}.

\section{Stabilizer codes and fault-tolerant flagged syndrome extraction}\label{sec.preliminaries}
  
 \subsection{Stabilizer Codes}

 We consider quantum computation with qubits using the computational basis $\{\ket{0}, \ket{1}\}$.
  The  Pauli matrices $I=\left(\begin{array}{cc}
 	1 & 0\\	0 & 1  
 \end{array}\right)$, $X=\left(\begin{array}{cc}
 	0 & 1\\
 	1 & 0
 \end{array}\right)$, $Y=\left(\begin{array}{cc}
 	0 & -i\\
 	i & 0
 \end{array}\right)$, and $Z=\left(\begin{array}{cc}
 	1 & 0\\
 	0 & {-1}
 \end{array}\right)$ form a basis for the linear operators on a single-qubit state space. 
 Similarly, the collection of the $n$-fold tensor product of the  Pauli matrices forms a basis for the linear operators on the $n$-qubit state space. 
For simplicity,  let $X_i$ denote the operator consisting of a Pauli matrix $X$ on the $i$-th qubit
and identities on the other qubits. $Z_i$ and $Y_i$ are  similarly defined.  Additionally, we may omit the tensor product symbol. Thus an $n$-fold Pauli operator can be represented by its nonidentity components. For instance,   $X\otimes Y\otimes Z\otimes I\otimes I$ will be represented as $X_1Y_2Z_3$.
Moreover, we may use the notation $X_T$, where $T$ is a set of positions, to indicate that an $X$ is applied to each qubit in $T$. 
For example, $X_{1,2,3}$ represents $X_1X_2X_3$.

The weight of an $n$-fold Pauli operator indicates the number of non-identity elements it contains. 
We consider quantum errors that are Pauli errors.   
Typically, errors with lower  weights are more likely than those with higher weights in an error model.

Let $\cS$ be an abelian subgroup of the $n$-fold Pauli operators with appropriate phases.
Suppose  $\cS$ has $n-k$ independent generators $g_1,\dots,g_{n-k}$.
Then it defines an $\codepar{5,1,3}$ stabilizer code $C(\cS)$,  which is   a $2^k$-dimensional subspace of the  $n$-qubit state space  that is fixed by $\cS$.
Thus $\cS$ is called a stabilizer group and the operators in $\cS$ are called stabilizers. Any non-identity $n$-fold Pauli operators of weight less than $d$ must either be in the stabilizer group  or anticommute with some stabilizers~\cite{Got97}.
Two Pauli operators either commute or anticommute with each other. 
The \textit{error syndrome} of a Pauli error $E$  with respect to the $n-k$ stabilizer generators $g_i\in\cS$ is a binary string of length $n-k$, whose $i$-th bit  is $1$ if $E$ anticommutes with $g_i$, and $0$, otherwise.
The error syndrome of an error describes its commutation relations with the stabilizer generators. 
 A code of distance $d$ is capable of correcting $\lfloor\frac{d-1}{2}\rfloor$ errors.

A Pauli operator consisting entirely of $X$ or $Z$ matrices is said to of $X$- or $Z$-type, respectively. 
A  CSS code~\cite{CS96,Ste96a} is a stabilizer code 
with some stabilizer generators of $X$-type and others of $Z$-type.

\subsection{Fault-tolerant quantum computation}
\label{sec:ftqc}

A quantum circuit comprises fundamental single-qubit and two-qubit quantum gates, along with state preparations and measurements. We will focus on qubit measurements in the $X$ or $Z$ basis, as well as the controlled-phase (CZ) gate  and controlled-NOT (CNOT) gate.

Errors in a quantum circuit can arise from  imperfect quantum gates, idle qubit, noisy qubit measurements, or other sources. Each of these sources will be referred to as a location.
However, the primary cause of a logical qubit error often stems from the error propagation through CZ and CNOT gates. 
Let $\mathrm{CNOT}_{1,2}= \ket{0}\bra{0}\otimes I+ \ket{1}\bra{1}\otimes X$,
where the first and second qubits are called the control and target qubits, respectively. 
Similarly, let $\mathrm{CZ}_{1,2}= \ket{00}\bra{00}+\ket{01}\bra{01}+\ket{10}\bra{10}-\ket{11}\bra{11}$.
Then we have $\mathrm{CNOT}_{1,2} X_1=  X_1X_2 \mathrm{CNOT}_{1,2}$,
$\mathrm{CNOT}_{1,2} Z_2=  Z_1Z_2 \mathrm{CNOT}_{1,2}$,
$\mathrm{CZ}_{1,2} X_1=  X_1Z_2 \mathrm{CZ}_{1,2}$, and
$\mathrm{CZ}_{1,2} X_2=  Z_1X_2 \mathrm{CZ}_{1,2}$.
As a result, a single-qubit Pauli error occurring in one of the qubits involved in a CNOT or CZ gate may propagate to the other qubit.

Each quantum gate, qubit state preparation,  or qubit measurement takes the same unit of time.  The gates and error model considered in this paper is defined as follows.
 \begin{enumerate}
 	\item Gate Errors: Independent depolarizing errors with rate $p$ are introduced after perfect quantum gates in the circuit.
 	\begin{itemize}
 		\item Following each CNOT or CZ gate, one of the 15 non-identity two-qubit Pauli operators is applied with probability $p/15$. 
 		\item Ancillary state preparations ($\ket{0},$ $\ket{1},$ $\ket{+},$ $\ket{-}$) are corrupted with  probability  $2p/3$.
 	\end{itemize}
 \item Idle Qubit Errors: idle qubits are subject to depolarizing errors with rate $\gamma p$, where $0\leq \gamma\leq 1$.
 \item Measurement Errors: Independent depolarizing errors with rate $\beta p$ are introduced before measurements,
 where $\beta$ could be larger than $1$.
 \end{enumerate}

A \textit{location failure} denotes  an event that a Pauli error occurs after a perfect single-qubit gate,
two-qubit gate, ancilla preparation, an idle qubit or before a  perfect qubit measurement in a circuit.
	We consider an FTQC scheme built upon a quantum code with distance $d$, aiming to correct any $t=\lfloor\frac{d-1}{2}\rfloor$ location errors in a round of error correction.
Specifically, we focus on the SE process in an FTQC scheme,  which involves the interaction of data qubits of the quantum codeword with additional ancillary qubits for syndrome measurements, using CNOT and CZ gates.

	If a set of location failures of size $w$
	occurs in a circuit, while the other locations are noiseless, this is referred to as a \textit{$w$-fault event}.
	 If the evolution of a $1$-fault event results in a Pauli error of weight higher than one on the data qubits,
	 this event is considered a \textit{high-weight error propagation event}.
	 A procedure  is  \textit{fault-tolerant}  if  a total of no more than $t$ location errors on the input quantum state and during the procedure  can be corrected by a perfect error correction procedure.

 A round of SE refers to a sequence of stages in a circuit designed to measure a complete set of stabilizer generators, where each stage is designed to extract the error syndrome bits of one or more stabilizers. Since imperfect gates are used for SE, the syndrome measurement outcomes are not reliable, and multiple rounds of SE are usually applied to ensure that reliable error information can be extracted with high probability.
 For a stabilizer code with  minimum distance $d$, a total of $L=O(d)$ rounds of SE is sufficient to achieve fault-tolerant SE~\cite{DKLP02,KL24}. The collection of all  measurement outcomes from more than one rounds of SE is referred to as a \textit{generalized error syndrome}.

To perform error correction based on a given generalized error syndrome, one would like to find the most probable error event matching the syndrome. One approach is to construct a \textit{lookup table} consisting of pairs of error syndromes and corresponding error operators for error recovery on the data qubits. 
 
 \bd
Two error events are said to be \textit{degenerate}  if they generate the same residual Pauli errors on the data qubits up to a stabilizer.	 
\ed
Consequently, there is no need to distinguish between degenerate error events.

\bp An SE procedure for a code with minimum distance $d$ is fault-tolerant if  any $w$-fault event, with $1\leq w\leq \lfloor\frac{d-1}{2}\rfloor$,  has a unique generalized error syndrome  up to degeneracy.  
\ep

Since single-qubit and two-qubit gates dominate a quantum circuit, we adopt the following definition of error threshold with respect to the gate error rate.

\bd \label{def:threshold}
The \textit{error threshold} of a scheme is the value at which, if the physical gate error rate is below this threshold, the logical error rate of the procedure will be lower than the physical gate error rate.
\ed

As a result, the error threshold for implementing a fault-tolerant quantum memory will be close to the threshold for implementing a fault-tolerant logical CNOT gate.

\subsection{Fault-tolerant syndrome extraction with flag qubits}

\begin{figure}[htbp]
	\centering
	\includegraphics[width=0.95\linewidth]{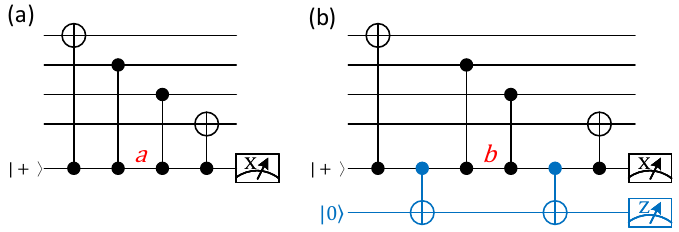}
	\caption{Syndrome extraction circuits for measurer a stabilizer $X_1Z_2Z_3X_4$: (a) a raw SE circuit; (b) a flagged syndrome extract circuit. The ancilla qubit initialized in $\ket{0}$  is a flag qubit. The blue part includes a pair of CNOT gates used to detect whether a high-weight error propagation event has occurred. }\label{fig.unflag_flag}
\end{figure}

An SE circuit is used to determine the error syndrome of an occurred error. 
Figure ~\ref{fig.unflag_flag} (a) illustrates a SE circuit for measuring  the Pauli operator $X_1Z_2Z_3X_4$ on four data qubits, using an ancillary qubit in $\ket{+}$.
 This ancillary qubit is called a \textit{measurement qubit} and the measurement outcome corresponds to the syndrome bit associated with  $X_1Z_2Z_3X_4$.
This circuit accurately extracts the syndrome bit under ideal conditions. 
However, when a Pauli $X$ error occurs at location $a$ in Fig.~\ref{fig.unflag_flag}\,(a), it propagates through a CZ and a CNOT gates, leaving a weight-2 error $Z_3X_4$ on the data qubits.
This constitutes a high-weight error propagation event that may require additional attention.

\begin{figure*}[hbtp]
	\centering
	\includegraphics[width=1\linewidth]{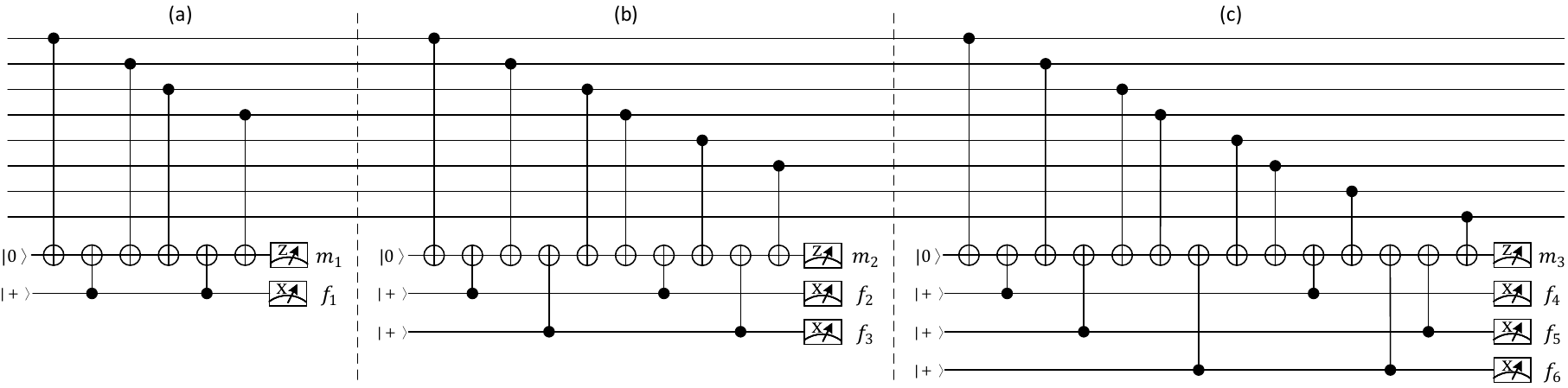}
	\caption{Parts (a), (b), and (c) are weight-4, weight-6, and weight-8  SE circuits, respectively, using one, two, and three flag qubits.   The measurement outcomes on the measurement qubits and flag qubits are denoted by $m_i$ and $f_j$, respectively.
	}\label{fig.2-flag_circuit3}
\end{figure*}

On the contrary, the circuit in Fig.~\ref{fig.unflag_flag}\,(b) introduces an ancillary qubit initialized to $\ket{0}$, referred to as a \textit{flag} qubit (highlighted in blue) includes an additional pair of CNOT gates. When a Pauli $X$ error occurs at location $b$, the measurement outcome on the flag qubit will be one, signaling a high-weight error propagation event \cite{CR18a,CR18b}. In this scenario, the occurrence of a high-weight error propagation event triggers the flag qubit, causing the flag to be raised. As a result, the circuit  in Fig.~\ref{fig.unflag_flag}\,(b) is  termed a \textit{flagged SE circuit}, while   the circuit in Fig.~\ref{fig.unflag_flag}\,(a) is termed an \textit{raw SE circuit}.

For stabilizers of higher weight, additional flag qubits are required for signaling high-weight error propagation events~\cite{CB18,CR20}.
Figure~\ref{fig.2-flag_circuit3} illustrates the SE circuits for $Z$-type stabilizers of weight $4, 6, $ and $8$, respectively,
using one, two, and three flag qubits~\cite{CB18}.
For stabilizers of higher weight, $O(w)$ CNOTs and flag qubits are required. A trivial flagged SE circuit can be constructed using  $O(w)$ flag qubits by connecting a CNOT gate from the measurement qubit to a flag qubit after each CNOT gate from the data qubit to the measurement qubit.
 An SE circuit for a weight-$w$ stabilizer   will be referred to as a weight-$w$ SE circuit.

 In conventional flagged SE schemes,  multiple rounds of SE are performed and a decision tree is constructed with a lookup table at each leaf (c.f. Fig.~\ref{fig.tree_d3} and Fig.~\ref{fig.tree_d5}). Routing on the decision tree depends on the outcomes of the measurement qubits and flag qubits. A lookup table is a precomputed one-to-one function that outputs a correction operator on input an error syndrome from a complete (raw) SE circuit. 
   Multiple lookup tables of potential errors must be constructed based on the flag outcomes. However, the outcomes on the measurement qubits  are omitted when some flag qubits rises in the same round of SE. Thus we have the  following proposition.

\begin{proposition}
	 In flagged SE, the outcomes of the measurement qubits and flag qubits   in $L$ rounds of flagged or raw SE  can be collected to infer the most likely residual error on the data qubits.  \label{prop:tag_measurement}
\end{proposition}
Using additional error information from the measurement qubits helps design an SE circuit with a unique generalized error syndrome for each target error event.
This technique is critical for our construction of flag-sharing schemes with shorter circuit depths later.
 
We defer the discussion of Proposition~\ref{prop:tag_measurement} to Subsec.~\ref{subsec:lookup_contruction}, where we illustrate its utility using the example of the $\codepar{5,1,3}$ code for clarity and simplicity.

 In flagged SE, $L=O(d)$ rounds of flagged or raw SE is sufficient for FT SE.
By carefully examining the error correction decision tree, it is possible to use fewer than 
$d$ rounds of SE on average as we will show in Section~\ref{sec:QEC}.

 \section{Versatile syndrome extraction schemes for CSS codes}
 \label{sec:SE_CSS}

 For a given independent set of stabilizer generators, the seminal flagged FTQC scheme adopts a serial implementation of SE, as illustrated in Fig.~\ref{fig.2-flag_circuit3}~\cite{CR18a,CR18b}, which reduces the number of ancillary qubits required for fault-tolerant SE.  
 It is possible to extract multiple syndromes simultaneously using shared flag qubits \cite{Rei20,LL23,DML+24}, thus reducing the depth of SE.
 
 To design a fault-tolerant SE scheme, it is important to minimize sources of errors and keep these errors under control to achieve a good error threshold. We define the notion of circuit area as follows:
 \bd
 	The effective circuit area of an SE circuit  for a quantum code is defined as:
 \[
2n_{\mathrm{g}}+n_{\mathrm{p}}+\beta n_{\mathrm{m}}+\gamma n_{\mathrm{i}},
\]
 	where $n_{\mathrm{g}}$, $n_{\mathrm{p}}$, $n_{\mathrm{m}}$, and $n_{\mathrm{i}}$ are the numbers of two-qubit gates, ancilla preparations, measurements, and idle qubits in $\mathcal{C}$, respectively.
  \ed
 Intuitively, an SE circuit  with a low effective circuit area  will have a better error threshold. When the idle qubit error rate is high, the error threshold degrades significantly. Therefore, achieving good parallelism in SE to minimize the number of idle qubits is desirable. Moreover, when the measurement error rate dominates, it is preferable to reduce the number of measurements by using fewer flag qubits in SE.

In this section, we propose versatile SE circuits to reduce the effective circuit area. These circuits are categorized into three types: (1) serial SE, (2) fully parallel SE, and (3) sequential flag-sharing SE.

Roughly speaking, a SE scheme is represented by an integer matrix $M$,
whose  $(i,j)$-th entry $M_{i,j}$ denotes that   $M_{i,j}$ stabilizers are simultaneously measured using the $i$-th set of flag qubits at stage $j$. The sum $s=\sum_{i,j} M_{i,j}$  equals the total number of stabilizers to be measured.  
By considering $M_{i,j}$ as an integer partition of $s$, versatile SE circuits can be designed.
(For example, in Fig.~\ref{fig:s2p}, (a) is denoted as a $[1~1]$ scheme while (b) is a $[2]$ scheme.)

To illustrate the concepts, we first use the $\codepar{17,1,5}$ code as an example, which is defined with stabilizers: 
\begin{equation}\label{eq:1715_stabilizers}
	\begin{aligned}
		g_1=&Z_1Z_2Z_3Z_4,\,g_2=Z_1Z_3Z_5Z_6,&\\
		g_3=&Z_5Z_6Z_9Z_{10},\,g_4=Z_7Z_8Z_{11}Z_{12},&\\
		g_5=&Z_9Z_{10}Z_{13}Z_{14},\,g_6=Z_{11}Z_{12}Z_{15}Z_{16},&\\
		g_7=&Z_8Z_{12}Z_{16}Z_{17},\,g_8=Z_3Z_4Z_6Z_7Z_{10}Z_{11}Z_{14}Z_{15},&\\
		g_9=&X_1X_2X_3X_4,\,g_{10}=X_1X_3X_5X_6,&\\
		g_{11}=&X_5X_6X_9X_{10},\,g_{12}=X_7X_8X_{11}X_{12}.&\\
		g_{13}=&X_9X_{10}X_{13}X_{14},\,g_{14}=X_{11}X_{12}X_{15}X_{16},&\\
		g_{15}=&X_8X_{12}X_{16}X_{17},\,g_{16}=X_3X_4X_6X_7X_{10}X_{11}X_{14}X_{15},&
	\end{aligned}
\end{equation}
and   logical operators:
\begin{equation}
	\bar{X}=X^{\otimes 17}, \:\bar{Z}=Z^{\otimes 17}.
\end{equation}
Since $g_1,\dots,g_8$ are of $Z$-type and $g_9,\dots,g_{16}$ are of $X$-type and they are equivalent up to a Hadamard transform,  we focus on the  design of SE for  $Z$-type stabilizers.

\begin{figure}[htbp]
	\centering
	\includegraphics[width=0.9\linewidth]{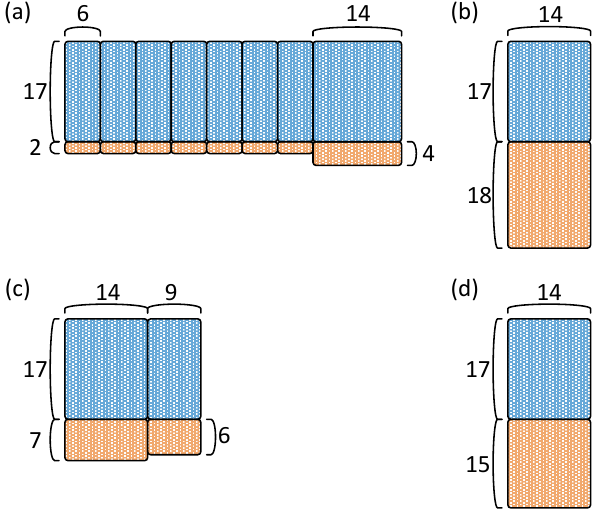}
	\caption{
			Various SE schemes for the \codepar{17,1,5} code are illustrated:
		(a) a serial [1~1~1~1~1~1~1~1] SE scheme;
		(b) a fully parallel [1;~1;~1;~1;~1;~1;~1;~1] scheme;
		(c) a  sequential [4~4] flag-sharing scheme;
		(d) a  sequential [2;~2;~2;~1;~1] flag-sharing scheme.
		 A pair of a blue box and an orange box denote a stage of SE. The height of the blue box represents the number of data qubits of the code, and the height of the orange region indicates the number of ancilla qubits required. The width of the blue box indicates the circuit depth.  	 
		 Measurements and state preparations are not shown in these illustrations. 
	}\label{fig.tile_1715}
\end{figure}

Four SE schemes for the $\codepar{17,1,5}$ code are presented in Fig.~\ref{fig.tile_1715}. 
We will explore these approaches in the following subsections.

\subsection{Serial syndrome extraction}  
 A traditional flag scheme  utilizes a serial SE circuit 
to measure all the stabilizer generators~\cite{CR18b,CB18,BSE+23},
which is referred to as a   $\mathbf{1}^T$ scheme.
This approach helps reduce the number of ancillary qubits and measurements.

For an $n$-qubit code with $s$ stabilizer generators of mean weight $w$, the number of flag qubits required for each stabilizer is roughly $w/2$ and a stage of SE circuit is of depth $w+ 2(w/2)=2w$. 
Consequently, the  $\mathbf{1}^T$ scheme  involves    roughly $2ws$ CNOT gates,  $(s+ws/2)$ state preparations, $(s+ws/2)$ measurements
and $s(2w(n+1+w/2)-4w)$ idle qubits.
Thus the effective circuit area is
approximately 
\begin{align}
\beta(s+wr/2)+(s+9ws/2)+\gamma (2wsn+sw^2-2ws). \label{eq:area_ser}
\end{align}

For the \codepar{17,1,5} code, seven of the $Z$-type stabilizers  have  weight 4, while the remaining one has weight 8. Thus we have a [1 1 1 1 1 1 1 1 ] scheme as shown in Fig.~\ref{fig.tile_1715}~(a),
where the first seven stages follow from Fig.~\ref{fig.2-flag_circuit3}~(a)
and the last stage is from Fig.~\ref{fig.2-flag_circuit3}~(c).

This scheme introduces many idle qubits, making it suitable for scenarios with a low idle qubit error rate.

\subsection{Fully parallel syndrome extraction}

In order to reduce the number of idle qubits, we propose a fully parallel SE scheme, denoted as a $\mathbf{1}$ scheme, which utilize the full set of flag qubits from a  serial $\mathbf{1}^T$ SE scheme  to decrease the circuit depth.  

\begin{figure}[htbp]
	\centering
	\includegraphics[width=0.99\linewidth]{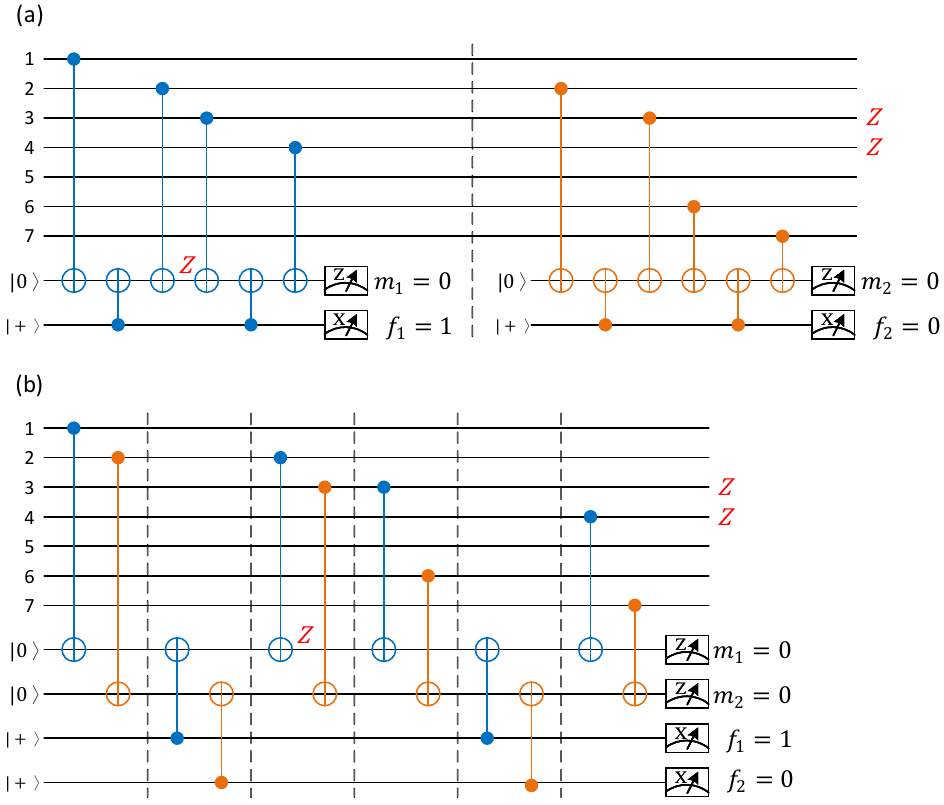}
	\caption{ SE circuits for two weight-4 $Z$-type stabilizers: (a) a serial scheme;  (b) a parallel scheme.	}
	\label{fig:s2p}
\end{figure}

Two stages of SE circuits in a   $\mathbf{1}^T$  scheme can be parallelized by interleaving their gates, enabling simultaneous execution of operations as shown in Fig.~\ref{fig:s2p}.

\bl  
Suppose  that a  quantum code allows for FT SE of $r$ $Z$-type (or $X$-type) stabilizers  using a serial flagged SE scheme  with $L$ rounds of SE. Then
the  $r$ stages of SE in the serial scheme can be parallelized by interleaving  the CNOTs in their order on the data qubits.

\label{lemma:s2p}
\el
\noindent\textbf{proof.}
Consider a code of distance $d$ and $t=\lfloor \frac{d-1}{2}\rfloor$.   

Let us consider each $1$-fault event in the $\mathbf{1}$ scheme, which has a corresponding $1$-fault event in the original $\mathbf{1}^T$ scheme. Such an example is shown in Fig.~\ref{fig:s2p}\,(a) and\,(b).

An important observation is that in the flagged SE circuit in Fig.~\ref{fig.2-flag_circuit3}, a state preparation or measurement fault will not propagate. Only the $1$-fault events leaving a $Z$ error on the measurement qubit will propagate back to the data qubits, triggering only the following $X$ measurement qubits but no flag qubits. Now, if we interleave the SE circuits as shown in  Fig.~\ref{fig:s2p}\,(b), we obtain a similar result where a state preparation or measurement fault will not propagate. Thus, a state preparation or measurement fault in the $\mathbf{1}$ scheme will have the same generalized error syndrome as its corresponding $1$-fault event in the original $\mathbf{1}^T$ scheme.

Now consider a $1$-fault high-weight error propagation event in the $\mathbf{1}$ scheme, which triggers some flag qubits and propagates some $Z$ errors into the data qubits. Again, these $Z$ errors on the data qubits will trigger only the following $X$ measurement qubits but no flag qubits. It can be observed that a corresponding $1$-fault high-weight error propagation event in the $\mathbf{1}^T$ scheme will trigger corresponding flag qubits and propagate the same $Z$ errors to the data qubits. Thus, these two $1$-fault high-weight error propagation events will have the same generalized error syndrome.
Such an example is shown in Fig.~\ref{fig:s2p}.

By the assumption that the $\mathbf{1}^T$ scheme is fault-tolerant, any $t$-fault event has a unique generalized error syndrome. Therefore, the $\mathbf{1}$ scheme is fault-tolerant.

\hfill $\square$

More generally, we have the following corollary.
\bc
Given an FT $\mathbf{1}^T$ SE scheme for $s$ $Z$-type (or $X$-type) stabilizers,
one can derive an $M$ SE scheme for any binary matrix $M$ such that $\sum_{i,j}M_{i,j} =s$. 
In particular, an FT  $\mathbf{1}^T$ SE scheme can be transformed into a  $\mathbf{1}$ SE scheme while maintaining fault tolerance. \label{cor:s2p}
 \ec
 
It is straightforward to see that a $\mathbf{1}$ scheme would have the lowest circuit depth and possibly the lowest effective circuit area among possible binary matrices $M$ with  $\sum_{i,j}M_{i,j} =s$. We will focus on this variation in the following discussion. 
 
 \begin{remark} \label{remark:cnot_order}
		The circuit depth of the 
 $\mathbf{1}$ scheme is not necessarily the depth of its longest stage. This is because the CNOT order on each data qubit must be maintained as in the original $\mathbf{1}^T$ scheme. One may reorder the CNOTs to reduce the circuit depth, but it is essential to ensure that any $n$-fault event has its unique generalized error syndrome.
 Consequently, it is important to include the outcomes of the measurement qubits as a part of the generalized error syndrome by Prop.~\ref{prop:tag_measurement}. (See Subsec.~\ref{subsec:lookup_contruction} for a detailed discussion on the \codepar{5,1,3} code.)
 \end{remark}

Consider $r$ stabilizers of mean weight $w$, with the highest weight being $2w$.
Assuming the circuit depth is $3w$, the effective circuit area  is about  $ \gamma (3w(n+r+ w/2)- 4rw )+4rw  + (1+\beta)(r+ rw/2)$. If $r=s/2$, the effective circuit area of two $\mathbf{1}$ SE schemes for $Z$- and $X$-type stabilizers is about
\begin{align}
\gamma(6wn +sw+3sw^2/2)+2sw+(1+\beta)(s+sw/2). \label{eq:area_par}
	\end{align}
For $\gamma=0$ and $\beta=1$, Eqs. (\ref{eq:area_par}) and (\ref{eq:area_ser}) are similar.
For $\gamma=\beta=1$, Eq. (\ref{eq:area_par}) has a dominant term $2 wsn$. Consequently,
a $\mathbf{1}$ scheme is more suitable than a $\mathbf{1}^T$ scheme if ancillary qubits are sufficient.

For the \codepar{17,1,5} code, we adopt the  $\mathbf{1}^T$ SE scheme for the $Z$-type stabilizers in Eq.~(\ref{eq:1715_stabilizers}) to construct a $\mathbf{1}$ SE scheme as shown in Fig.~\ref{fig.tile_1715} (b).
By~Lemma~\ref{lemma:s2p}, we can directly parallelize the  $\mathbf{1}^T$ SE scheme  to generate the  $\mathbf{1}$ SE scheme. The detailed circuit of Fig.~\ref{fig.tile_1715} (b) is depicted in  Fig~\ref{fig.1715_full} in Appendix~\ref{app:1715}, where the circuit depth is $14$, the same as the length of the longest stage in the $\mathbf{1}^T$ scheme.
The flagged SE circuit for the weight-8 $Z$-type stabilizer dominates the circuit depth, which is 14 according to Fig.~\ref{fig.2-flag_circuit3}.
There are ten flag qubits in total, of which seven are for the seven weight-4 stabilizers.

 \subsection{Sequential flag-sharing syndrome extraction}

       This section will introduce how parallelism in SE can be achieved using versatile flag-sharing techniques, generalizing the ideas in~\cite{Rei20,LL23,DML+24},  to reduce the circuit depth while conserving the number of flag qubits.

\begin{figure}[htbp]
        	\centering
        	\includegraphics[width=1\linewidth]{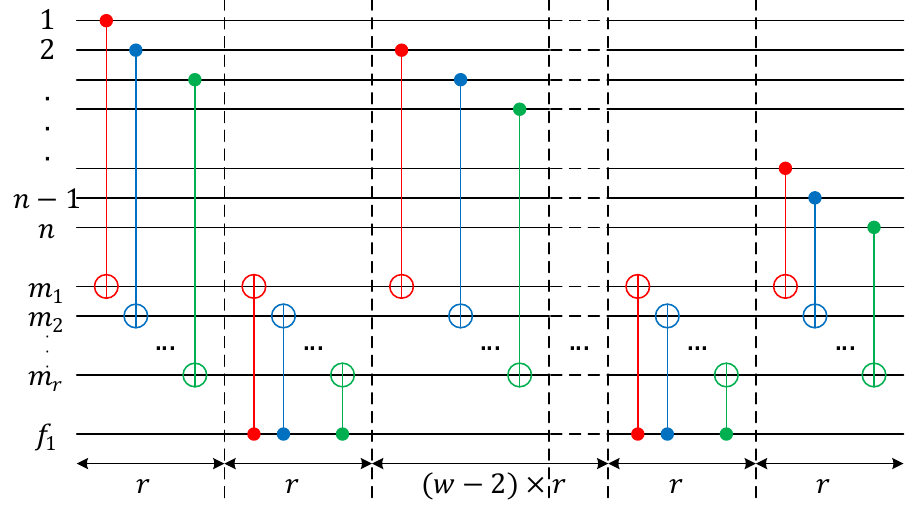}
        	\caption{ 		
        		A parallel flagged SE circuit with $r$ stabilizers sharing one flag qubit.  
        		This circuit extracts the syndrome $g_1$ (red), $g_2$ (blue), $\ldots$, and $g_r$ (green).
        		It is capable of detecting weight-1 error events through a pair of CNOT operations between measurement qubits and the flag qubit. The solid black lines below the diagram indicate the total number of CNOT gates used within this interval. Ancillary state preparations and measurements have been omitted for clarity.
        	} \label{fig.parallel_flag_d3}
        \end{figure}

       Similar to the discussion in Lemma~\ref{lemma:s2p}, we can merge multiple stages of SE circuits in a serial scheme by flag-sharing  to achieve parallelism as well.
       A parallel flag-sharing SE circuit using one flag qubit for distance-3 codes is shown  in Fig.~\ref{fig.parallel_flag_d3}. However, since fewer flag qubits are used, these stages can be merged only if any $t$-fault event has its unique generalized error syndrome up to degeneracy. Thus, the number of stages that can be merged is restricted in a flag-sharing scheme.
        
         Consider an \codepar{n,k,d} code with $(n-k)/2$ $Z$-type stabilizers and  $(n-k)/2$ $X$-type stabilizers.
Suppose that there is a serial SE scheme of $s=(n-k)/2$ stages  for the $Z$-type stabilizers $g_1,\dots,g_{s}$.
Assume that $g_i$ is of weight $w_i$ and  at stage $i$, $v_i$ flag qubits and additional $u_i$ CNOTs are used for flagged SE.

We would like to merge  $r$ stages of SE, say $g_1,\dots,g_r$, by sharing a set of flag qubit, as illustrated in Fig.~\ref{fig.parallel_flag_d3}.
Note that high-weight error propagation events occur only after the first $r$ CNOTs and before the last $r$ CNOTs that connect the measurement qubits and flag qubits. 
Since  there are $ \sum_{i=1}^r (u_i +w_i)$ CNOTs, we have at most $\Big(  \sum_{i=1}^r (u_i +w_i)-2r\Big)$ $1$-fault locations   that may induce high-weight error propagation events.

 The error information we can learn from this SE scheme is  the $r$-bit  measurement outcomes in this circuit and another $s$-bit measurement outcomes from the next complete SE circuit for $s$  $X$ stabilizers  according to Prop.~\ref{prop:tag_measurement}. 
To ensure that any combination of $t$ high-weight error propagation events has a unique generalized error syndrome for error recovery,
we have the following design criterion

\bp
 In a serial SE scheme,  $r$ stages  may be parallelized by sharing a set of flag qubits if \begin{align}
 	2^{s+r+\max_i v_i} \geq  8^t {\sum_{i=1}^r (u_i +w_i)-2r \choose t }.
 \end{align}
\ep
             This entire circuit will be denoted as an $[r]$ SE scheme in the matrix representation.
            The factor $8^t$ is because a CNOT gate error with a $Z$ or $Y$ error on the ancilla qubit  may cause a high-weight error propagation event.  
        Note that the order of CNOTs needs to be adjusted to ensure the establishment of a unique generalized error syndrome for each $1$-fault event.

      More generally, it is possible to design a sequential $[r_1~r_2~\cdots~r_\ell]$ SE scheme for $s$  stabilizers of $Z$-type,
      where $r_i$ stabilizers are measured using a set of flag qubits at stage $i$ and $\sum_{i} r_i = s$.
      We must ensure any $t$-fault event in a round of SE has its unique generalized error syndrome.
Following a similar idea, a $[4~4]$ SE scheme for the $\codepar{17,1,5}$ was proposed in~\cite{DML+24}, as shown in~Fig.~\ref{fig.tile_1715}\,(c).

Similar to the proof of Lemma~\ref{lemma:s2p}, we have the following lemma that transforms a general sequential flag-sharing SE scheme into a parallel flag-sharing SE scheme. 
\bl  
Suppose  that a  quantum code allows for FT SE of $r$ $Z$-type (or $X$-type) stabilizers  using a serial $[r_1~r_2~\cdots~r_\ell]$ SE scheme  with $L$ rounds of SE. Then this  $[r_1~r_2~\cdots~r_v]$  SE scheme can be transformed into a  $[r_1;~r_2;~\cdots;~r_\ell]$  SE scheme while maintaining fault tolerance.

\label{lemma:s2p2}
\el

\begin{remark}	
	Again, the circuit depth of the $[r_1;~r_2;~\cdots;~r_\ell]$  scheme is not necessarily the depth of its longest stage due to the constraint on the CNOT order on each data qubit, as mentioned in Remark~\ref{remark:cnot_order}. One may try to reorder the CNOTs to reduce the circuit depth while ensuring that any $n$-fault event has its unique generalized error syndrome.
\end{remark}

According to Lemma~\ref{lemma:s2p2}, we first construct a $[2~2~2~1~1]$ SE scheme for the $Z$-type stabilizers of the $\codepar{17,1,5}$ code and then transform it into   a $[2;~2;~2;~1;~1]$ SE scheme by Prop.~\ref{prop:tag_measurement}, as shown in Fig.~\ref{fig.tile_1715}\,(d).  This scheme uses a total of 15 ancilla qubits and the circuit depth is 14. Moreover, it has the smallest effective area among the SE schemes in~Fig.~\ref{fig.tile_1715}. The details are provided in~Fig.\,\ref{fig.1715_parallel} in Appendix~\ref{app:1715}.

\section{Versatile syndrome extraction schemes for non-CSS codes}
\label{sec:SE_nonCSS}

In the previous section, we studied various SE schemes for CSS codes. Extending those results to non-CSS codes is more challenging since an SE stage of a stabilizer may involve the interleaving of multiple CNOT and CZ gates, leading to complex error propagation events. For instance, an error on the measurement qubit error might propagate to a data qubit through a CZ (or CNOT) gate and then propagate back to another measurement qubit through a CNOT (or CZ) gate.

\begin{figure}[htbp]
	\centering
	\includegraphics[width=0.9\linewidth]{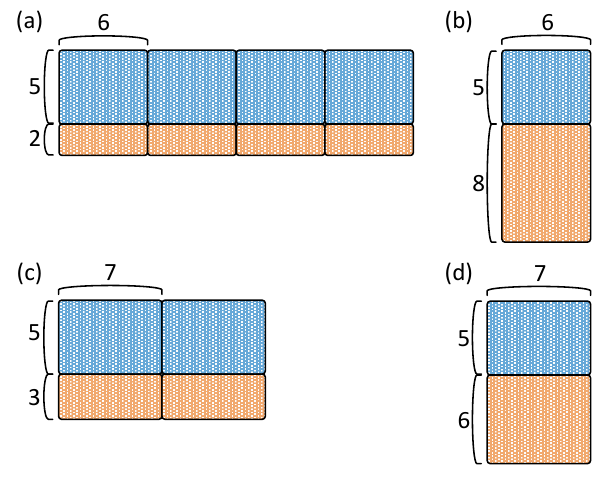}
	\caption{ 
		Various SE schemes for the \codepar{5,1,3} code are illustrated:
		(a) a serial [1~1~1~1] SE scheme;
		(b) a fully parallel [1;~1;~1;~1] scheme; 
		(c) a   [2~2] flag-sharing scheme;
		(d) a   $\begin{bmatrix}
			2\\2
		\end{bmatrix}$  flag-sharing scheme.
	}\label{fig.tile_513}
\end{figure}

\subsection{SE schemes for the \codepar{5,1,3} code}
Nonetheless, we propose versatile SE schemes for the $\codepar{5,1,3}$ code~{\cite{LMPZ96}} as shown in Fig.~\ref{fig.tile_513}. We consider the following five stabilizers
\begin{equation}\label{eq.513_stabilizers}
	\begin{aligned}
		g_1=&X_1Z_2Z_3X_4,\,g_3=X_1X_3Z_4Z_5,&\\
		g_2=&X_2Z_3Z_4X_5,\,g_4=Z_1X_2X_4Z_5.&\\
		g_5=&Z_1Z_2X_3X_5.&\\
	\end{aligned}
\end{equation}
Measuring any four of them one by one,  as shown in Fig.~\ref{fig.tile_513}\,(a),
constitutes  a [1~1~1~1] SE scheme. Each flag SE circuit stage is of depth 6 and uses two ancilla qubits.

\begin{figure}[htbp]
 	\centering
 	\includegraphics[width=0.9\columnwidth]{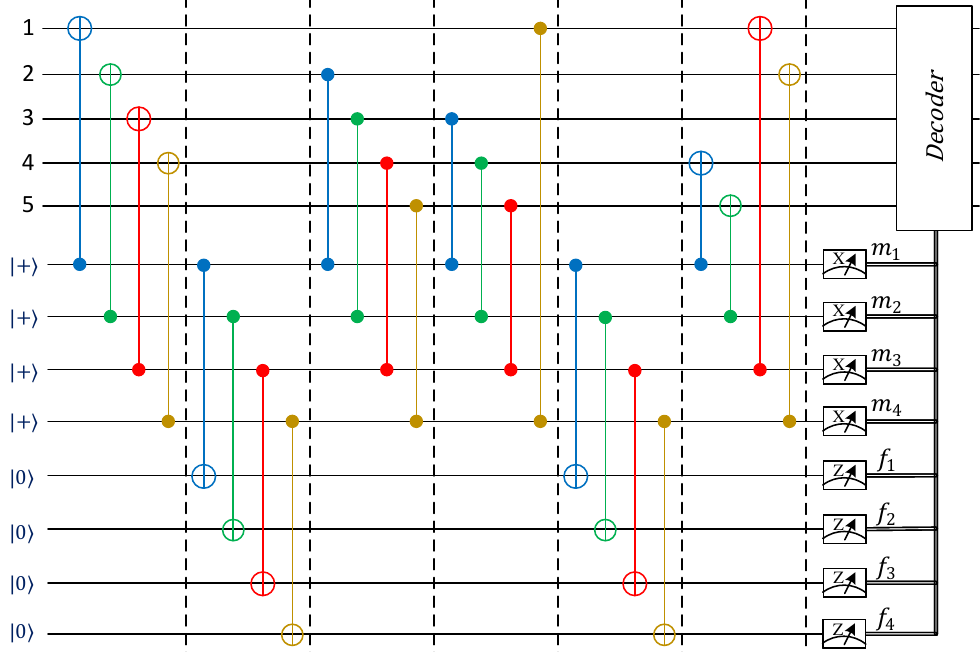}
 	\caption{The fully parallel $[1;~1;~1;~1]$ SE scheme for the \codepar{5,1,3} code, which extracts the syndrome bits of $g_1,g_2,g_3,$ and $g_4$.  Each color  marks an SE circuit of a stabilizer.}\label{fig.513_full_parallel_v1}
 \end{figure}

Similar to Cor.~\ref{cor:s2p},    we   can construct a fully parallel [1; 1; 1; 1] scheme for the \codepar{5,1,3} code,
as shown in Fig.~\ref{fig.tile_513}\,(b), and the details shown in~Fig.~\ref{fig.513_full_parallel_v1}.
This [1; 1; 1; 1] scheme has the smallest circuit depth of 6,
but uses a total $8$ ancilla qubits, of which $4$ are flag qubits.
A key observation is that the \codepar{5,1,3} code is a cyclic code, so we can manage to have all CNOTs or all CZs at each step of the circuit. In addition , a high-weight error propagation event will not propagate further
 
\begin{figure}[htbp]
	\centering
	\includegraphics[width=1\linewidth]{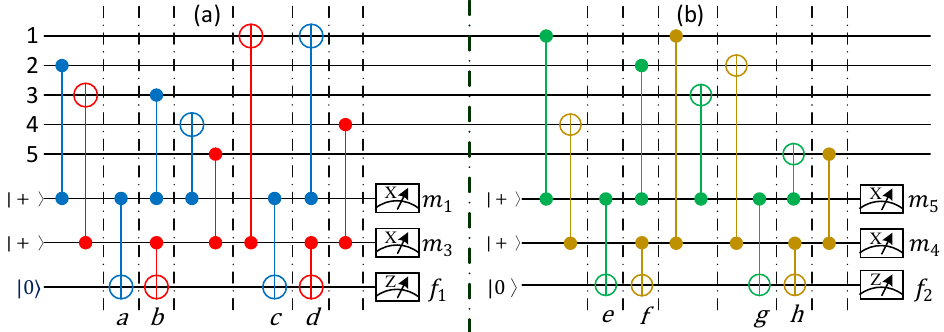}
	\caption{The [2~2] flag-sharing SE scheme for the \codepar{5,1,3} code, which extracts the syndrome bits of $g_1$ (blue) and $g_3$ (red) in part (a), and  $g_5$ (green) and $g_4$ (yellow) in the part (b). 
		CNOTs labeled from $a$ to $h$ are paired to couple a measurement qubit and the flag qubit, while sharing the same flag qubit. 
	}
	\label{fig.513_serial_parallel_v3}
\end{figure}

\begin{figure}[htbp]
	\centering
	\includegraphics[width=1\linewidth]{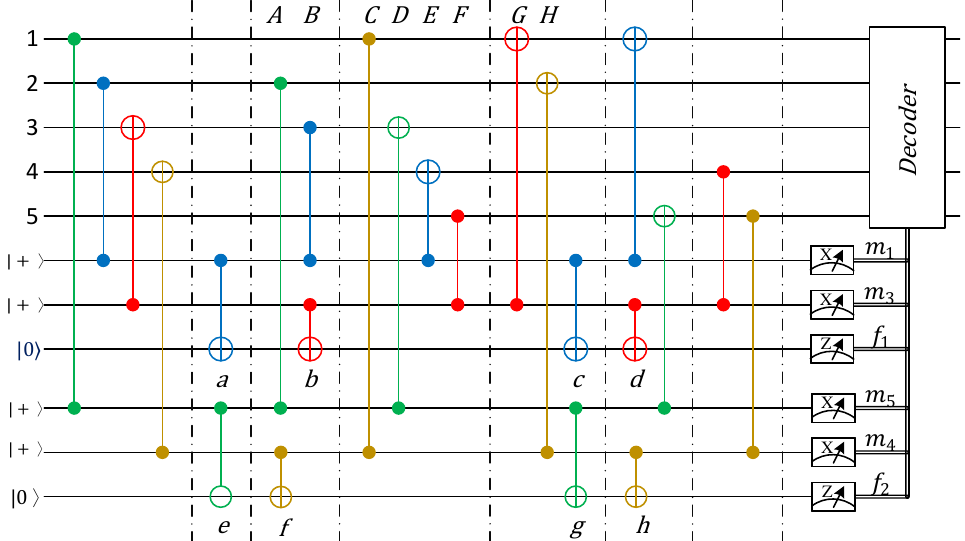}
	\caption{The $\begin{bmatrix}
			2\\2
		\end{bmatrix}$ SE scheme for the \codepar{5,1,3} code,  which extracts  the syndrome bits of $g_1$ (blue), $g_3$ (red), $g_5$ (green), and $g_4$ (yellow) using two flag qubits.  
	}
	\label{fig.513_parallel_parallel_v3}
\end{figure}

Building upon our previously proposed flag-sharing scheme~\cite{LL23}, 
two stabilizers can be simultaneously measured using three ancilla qubits, with one being a shared flag qubit,
resulting in a [2~2] structure as shown in Fig.~\ref{fig.tile_513}\,(c),  and the details shown in~Fig.~\ref{fig.513_serial_parallel_v3}.
While the depth of each SE circuit stage increases to 7, 
the total circuit depth decreases to 14, compared to the circuit depth of 24 in the serial [1~1~1~1] scheme.

Following the idea of  Lemma~\ref{lemma:s2p2}, we can transform the $[2~2]$ SE scheme  into a $\begin{bmatrix}
	2\\2
\end{bmatrix}$ SE scheme as shown in Fig.~\ref{fig.tile_513}\,(d), where two sets of shared flag SE circuits are used 
to  concurrently measure four stabilizers. The details of the scheme are provided in   Fig.~\ref{fig.513_parallel_parallel_v3},  which has the smallest effective circuit area among the four schemes in Fig.~\ref{fig.tile_513}, using only 6 ancillary qubits and 6 measurements.

We remark that  $g_1$, $g_5$, $g_3$,  and $g_4$ are measured in the  $[2~2]$ and $\begin{bmatrix}
	2\\2
\end{bmatrix}$ SE schemes.
By leveraging the outcomes of the measurement qubits when a flag rises and using the subsequent round of complete SE by Prop.~\ref{prop:tag_measurement}, we are able to distinguish a broader range of errors and construct legal lookup tables. Consequently,  we can construct  the $\begin{bmatrix}
	2\\2
\end{bmatrix}$ scheme with circuit depth $7$,
resulting in an effective circuit area lower than the fully parallel [1;~1;~1;~1] scheme.
However, if $g_1$, $g_2$, $g_3$, and $g_4$ are measured instead, we only find a  $\begin{bmatrix}
	2\\2
\end{bmatrix}$ scheme with circuit depth\,$8$.
The construction of  the lookup table for the $\begin{bmatrix}
	2\\2
\end{bmatrix}$ SE scheme will be discussed in Sec.~\ref{subsec:lookup_contruction} after we introduce the QEC procedure for the \codepar{5,1,3} code.

\section{Quantum error correction with flag qubits } 
\label{sec:QEC}

 This section provides QEC procedures for codes with distances 3 or 5, with potential extensions for codes with higher distances.

\subsection{Decision tree for distance-3 codes}\label{sec.decoder_d3}

 The flagged QEC rules for a distance-three quantum code are as follows. Initially, a complete flagged  SE circuit is implemented with outcomes on the measurement qubits $m$ and on the flag qubits $f$. If all measurement outcomes are zeros, the QEC procedure stops. Otherwise, nonzero measurement outcomes indicate the presence of a faulty event, and a complete raw SE circuit is performed, yielding outcomes $m'$.
 Error recovery is then performed based on all measurement outcomes $m,f,m'$ using a predetermined lookup table. 
 
 If no flag qubits are raised $(f=0)$, it suggests there are no high-weight error propagation events, and standard syndrome decoding proceeds based on the measured error syndrome from the complete raw SE circuit. However, if any flag qubit is raised, a high-weight error correction is chosen. This is done by exhaustively enumerating the 1-fault events and ensuring they have distinct generalized error syndromes up to degeneracy.
 Therefore, we can construct a lookup table indexed by $m$, $f$, and $m'$.

\begin{figure}[htbp]
 	\centering
 	\includegraphics[width=0.95\linewidth]{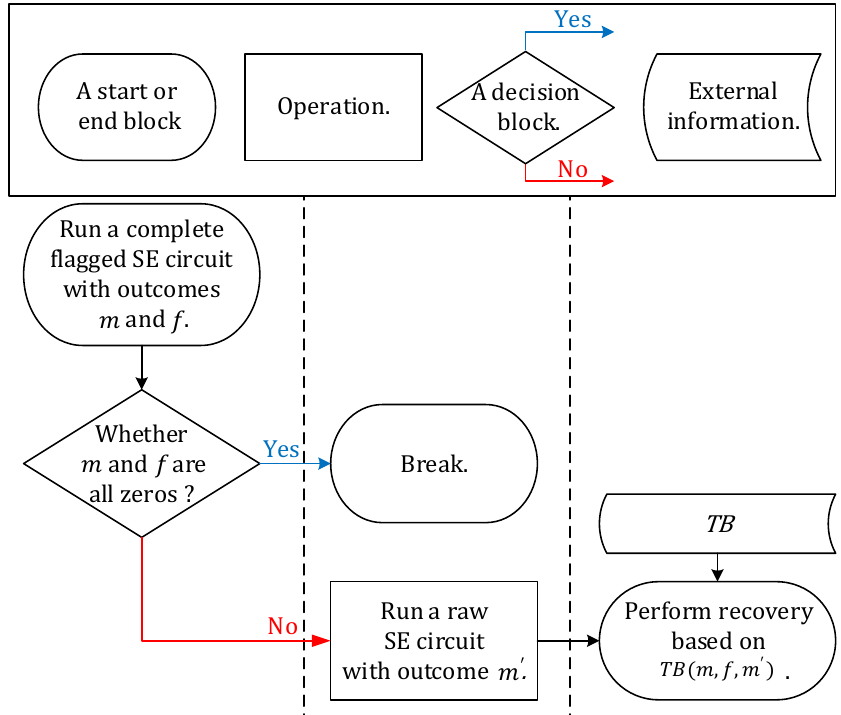}
 	\caption{The decision tree for distance-3 codes.
 		Each building block is listed in the top box. A rounded square represents the start or end of a procedure. A solid square represents an operation. A diamond indicates a decision with two outcomes, where the red line denotes the ``no" event, and the blue line denotes the ``yes" event. The last shape represents external data, such as the precomputed lookup table.
}\label{fig.tree_d3}
 \end{figure}

Figure~\ref{fig.tree_d3} depicts the decision tree for distance-3 codes.

\subsection{Constructing the lookup table for the [2;~2] SE scheme for the \codepar{5,1,3} code}~\label{subsec:lookup_contruction}

In this subsection, we demonstrate how to construct lookup tables for flagged SE schemes, using the $\begin{bmatrix}
	2\\2
\end{bmatrix}$ scheme for the \codepar{5,1,3} code as an example.  We will illustrate the effects of generalized error syndromes by utilizing the outcomes of measurement qubits and flag qubits as suggested by Prop.~\ref{prop:tag_measurement}.

Let us focus on the $\begin{bmatrix}
	2\\2
\end{bmatrix}$ SE scheme in Fig.~\ref{fig.513_parallel_parallel_v3}.

Let $m_{1534}=m_1m_5m_3m_4\in\{0,1\}^4$ and $f=f_1f_2\in\{0,1\}^2$ be the measurement outcomes on the measurement qubits and flag qubits in a  $\begin{bmatrix}
	2\\2
\end{bmatrix}$ SE circuit.
Let $m_{1534}'=m_1'm_5'm_3'm_4'\in\{0,1\}^4$  be the   measurement outcomes  of a complete raw  SE circuits of stabilizers $g_1,g_5,g_3,g_4$.

Consider Fig.~\ref{fig.513_parallel_parallel_v3}, the CNOT and CZ gates labeled by the uppercase letters A to H and lowercase letters a to h indicate possible locations, where $1$-fault events may  trigger flag qubits and generate high-weight error propagation events.
Since the SE circuit for the \codepar{5,1,3} code has both CNOT and CZ gates, both $X$ and $Z$ errors will propagate through the circuit. Thus, we need to account for the effects of $Y$ errors. We consider the following errors (
which are applied to the qubits in the order they appear in the circuit, from top to bottom):
\begin{itemize}
	\item A to H: $IX$, $XX$, $ZX$, $YX$, $IY$, $XY$, $ZY$, $YY$;
	\item a, b, e, f: $XI$, $XX$, $YI$,  $YX$;
	\item c, d, g, h: $XZ$, $XY$, $YZ$, $YY$. 
\end{itemize}
We have identified a total of 96 high-weight error propagation events.
For the lookup table to be effective, each high-weight error propagation event needs to have a unique syndrome, considering degeneracy.
A conventional flagged QEC (cf. \cite[Algorithm~1]{LL23}) using the outcomes of the flag qubits ($f$) and  a complete raw SE ($m'$) would generate at most $2^{2+4}$ distinct syndromes, which is insufficient.

By leveraging Prop.~\ref{prop:tag_measurement}, we can construct a lookup table of $2^{10}$ syndromes  using $m$, $f$, and $m'$.
Lookup tables for these high weight error propagation events are provided in Appendix~\ref{sec.appendix.513}

\be
In Fig.~\ref{fig.513_parallel_parallel_v3}, consider the (green) CZ at location A, which produces a Pauli error $XY$, denoted as $A_{XY}$.
This results in a high-weight error $X_{2,3,5}$, triggering  $f_2$.
Using a complete raw SE circuit, the outcomes would be 0111.

Next, consider an error $C_{YX}$ following the (yellow) CZ at location C, resulting in a high-weight error $Y_1X_2Z_{5}$, triggering $f_2$.
Again, a complete raw SE circuit generates  the syndrome bits 0111.

It can be observed that these two distinct high-weight error propagation events lead to the same syndrome.
Consequently, a standard lookup table cannot differentiate between these two errors.
However, leveraging the information from the  first round of SE circuit allows us to expand the lookup table.
For $A_{XY}$, $m_{1534}f_{12}$ is 010101, while for $C_{YX}$, it is 101001.
Therefore, the generalized error syndrome for $A_{XY}$ is updated to 0101010111,  and for $C_{YX}$, it is updated to 1010010111.
This enables the distinction between these two high-weight error propagation events for error correction purposes.

\hfill$\square$ 
\ee

Therefore, by utilizing  the expanded syndrome space and optimizing the CNOT order, we obtain a FT $\begin{bmatrix}
	2\\2
\end{bmatrix}$ SE scheme of depth $7$ for the \codepar{5,1,3} code.

\begin{figure*}[htbp]
	\centering
	\includegraphics[width=1\linewidth]{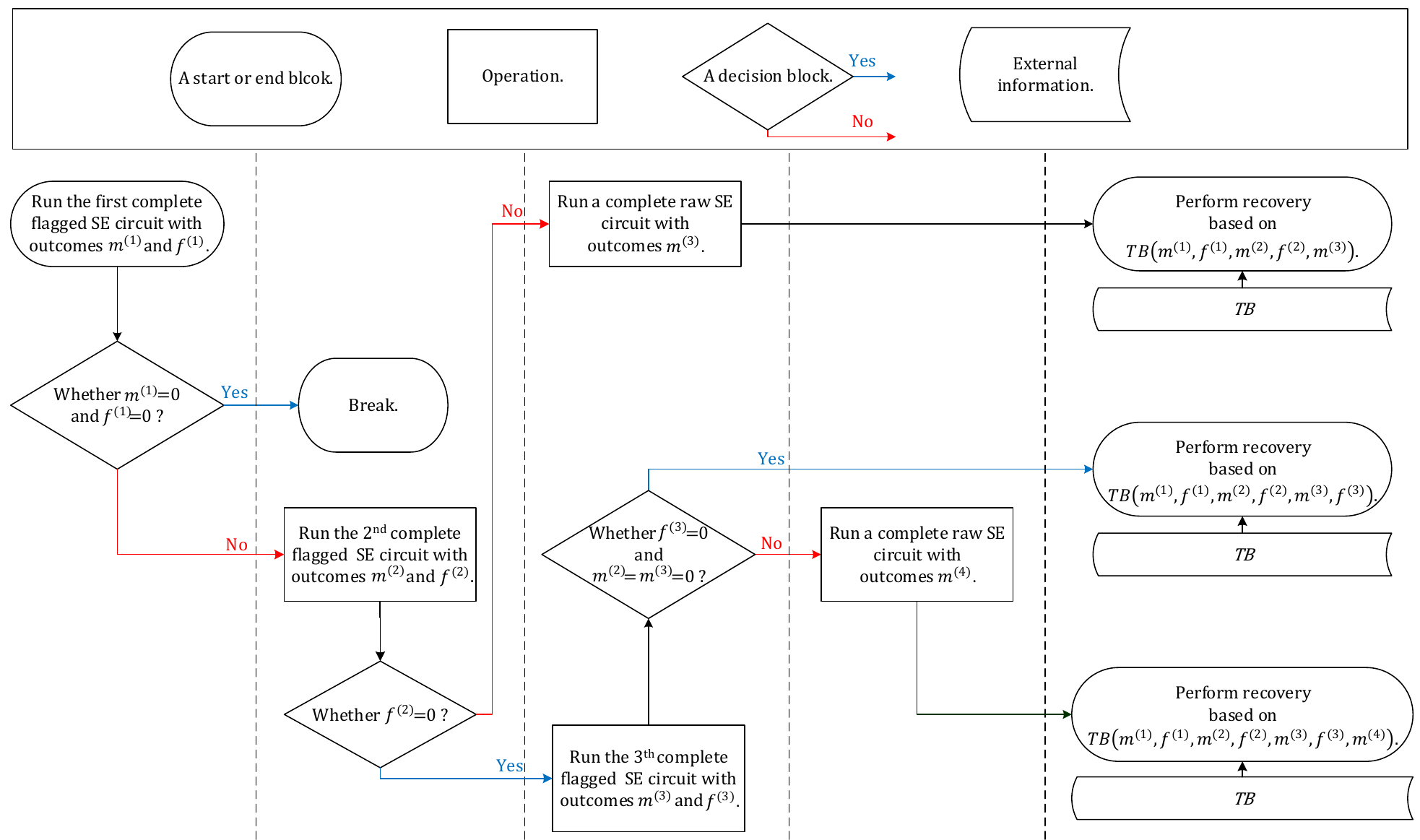}
	\caption{
		The decision tree for distance-5 codes.
		The building blocks are defined similarly to those in Fig.~\ref{fig.tree_d3}.
	}\label{fig.tree_d5}
\end{figure*}

\subsection{Decision tree for distance-5 codes}\label{sec.decoder_5}

The decision tree for distance-5 codes is depicted in Fig.~\ref{fig.tree_d5},
which extends the decision tree for distance-3 codes. Our goal is to correct any $w$-fault events for $w\leq 2$, using fewer than five rounds of flagged or raw SE. Therefore, we need to accurately identify no-fault, 1-fault, and 2-fault events. The decision tree has four routes, leading to four possible end blocks. We will explain each route as follows. Assume a lookup table, denoted  $TB$, has been precomputed.

If all the measurement outcomes of the first complete flagged SE circuit are zeros, we are confident that no faults have occurred. This scenario corresponds to the first end block in the decision tree.

When a single fault occurs in the first complete flagged SE circuit,  nonzero measurement outcomes $m^{(1)}$ and $f^{(1)}$ are obtained from the measurement and flag qubits, respectively. 
 We then run the second complete flagged SE circuit, which yields measurement outcomes $m^{(2)}$ and $f^{(2)}$. If $f^{(2)}$ is nonzero, it indicates a second fault that has triggered certain flag qubits, confirming a 2-fault event. Consequently, a complete raw SE is applied, yielding outcomes $m^{(3)}$. Recovery is then performed using the precomputed lookup table $TB$ with the entry $TB(m^{(1)}, f^{(1)}, m^{(2)}, f^{(2)}, m^{(3)})$. This scenario corresponds to the second end block in the decision tree. However, if the raw SE circuit also experiences a fault, it constitutes a 3-fault event, which exceeds the error correction capability of this code.

 On the other hand, if $f^{(2)}=0$, we may have either a 2-fault event or a 1-fault event. In either case, we need to run the third complete flagged SE circuit, resulting in measurement outcomes $m^{(3)}$ and $f^{(3)}$. If $f^{(3)}=0$ and $m^{(2)}=m^{(3)}$, we can confidently identify this as a 1-fault event and perform error recovery accordingly. Otherwise, if the conditions do not match, it confirms a 2-fault event, necessitating another round of complete raw SE to obtain outcome $m^{(4)}$ for error recovery. These scenarios correspond to the last two end blocks in the decision tree.

 Note that all measurement outcomes will be used to expand the syndrome space. Overall, we must construct a lookup table $TB$ indexed by $m^{(1)}, f^{(1)}, m^{(2)}, f^{(2)}, m^{(3)}, f^{(3)},$ and $m^{(4)}$ by exhaustively enumerating 2-fault events in the above procedure and ensuring they have distinct generalized error syndromes up to degeneracy.

\section{Simulation Results}\label{sec.simulation}

\begin{table*}[htbp] 
	\begin{tabular}{|c|c|c|c||c|c|c||c|c|c|}
		\hline
		Code & \multicolumn{3}{c||}{ \codepar{5,1,3}  } & \multicolumn{3}{c||}{ \codepar{17,1,5}  } & \multicolumn{3}{c|}{ \codepar{19,1,5}   }   \\
		\hline		                   
		Scheme         & \mbox{$\mathbf{1}^T$} & \mbox{$\begin{bmatrix}
				2\\2
			\end{bmatrix}$} 
		&  \mbox{$\mathbf{1}$} & \mbox{$\mathbf{1}^T$} &  \mbox{$\begin{bmatrix}		2\\2\\2\\1\\1
			\end{bmatrix}$} &  \mbox{$\mathbf{1}$}  &   \mbox{$\mathbf{1}^T$} &   \mbox{$\begin{bmatrix}		2\\2\\2\\1\\1\\1
			\end{bmatrix}$} &  \mbox{$\mathbf{1}$} \\ 
		\hline
		ancillary state preparation                  & 8      & 6    & 8    &18   &  15  &18   &21   & 18   &21 \\	
		measurement in $X$		               		 & 4      & 4    & 4    &10   &  7   &10   &12   & 9    &12\\	
		measurement in $Z$		 	           		 & 4      & 2    & 4    &8    &  8   &8    &9    & 9    &9\\		
		\hline
		Two-qubit gate         					     & 24     & 24   & 24   &56   &  56  & 56  &66   & 66   &66\\	
		Idle qubit 			   	                     & 120    & 29   & 30   &1064 &  336 & 378 &1320 & 238  &268\\	
		\hline \hline
		Circuit depth                                & 24      & 7    & 6   &56 & 14 &14 &66 & 10 &10\\	
		\hline 
		Physic qubits\footnote{($a_1$)+($a_2$): ($a_1$) represents the number of data qubits, and ($a_2$) represents the total number of measurement qubits and flag qubits.}                                            & 5+2& 5+6 & 5+8      &17+4 & 17+15 &17+18 &19+3 & 19+18 &19+21\\                                                  & 4 & 2 & 4      &10 &7 &10 &12 &9 &12\\
		\hline
		Circuit area ($\beta=1,\gamma=1$)  			 &184   &89       &94    &1212   &478    &526     &1494   &406    &442  \\
		Circuit area ($\beta=1,\gamma=0.01$)  		 &65.2  &60.29    &64.3  &158.64 &145.36 &151.78  &187.2  &170.38 &176.68  \\
		Circuit area ($\beta=10,\gamma=1$)  		 &256   &143      &166   &1374   &613    &688     &1683   &568    &631  \\
		Circuit area ($\beta=10,\gamma=0.01$)  	             &137.2 &114.29   &136.3 &320.64 &280.36 &313.78  &376.2  &332.38 &365.68  \\
		\hline
	\end{tabular}
	\caption{Resources required for a round of flagged SE for various schemes.
	}\label{table.list_area}
	
\end{table*}

In this section, we simulate various proposed SE schemes and compare these results  with the literature.
We also include  the  \codepar{19,1,5}  code~\cite{BM06}, which is defined by the stabilizers 
%\begin{equation}\label{eq:1915_stabilizers}
	\begin{align*}
		g_1=&Z_1Z_2Z_3Z_4,\,g_2=Z_1Z_3Z_5Z_7,&\\
		g_3=&Z_{12}Z_{13}Z_{14}Z_{15},\,g_4=Z_1Z_2Z_5Z_6Z_8Z_9,&\\
		g_5=&Z_6Z_9Z_{16}Z_{19},\,g_6=Z_{16}Z_{17}Z_{18}Z_{19},&\\
		g_7=&Z_{10}Z_{11}Z_{12}Z_{15},\,g_8=Z_8Z_9Z_{10}Z_{11}Z_{16}Z_{17},&\\
		g_9=&Z_5Z_7Z_8Z_{11}Z_{12}Z_{13},&
			\end{align*}
			\begin{align*}
		g_{10}=&X_1X_2X_3X_4,\,g_{11}=X_1X_3X_5X_7,&\\
		g_{12}=&X_{12}X_{13}X_{14}X_{15},\,g_{13}=X_1X_2X_5X_6X_8X_9,&\\
		g_{14}=&X_6X_9X_{16}X_{19},\,g_{15}=X_{16}X_{17}X_{18}X_{19},&\\
		g_{16}=&X_{10}X_{11}X_{12}X_{15},\,g_{17}=X_8X_9X_{10}X_{11}X_{16}X_{17},&\\
		g_{18}=&X_5X_7X_8X_{11}X_{16}X_{17},&
	\end{align*}
%\end{equation}
and the logical operators are
\begin{equation}
	\bar{X}=X^{\otimes 19}, \:\bar{Z}=Z^{\otimes 19}.\notag
\end{equation}
Various SE schemes for the \codepar{19,1,5} \cite{BM06} are illustrated in Fig.~\ref{fig.tile_1915},
 with detailed explanations provided in  Appendix~\ref{app:1915}.

We consider the circuit-level noise model as outlined in Sec.~\ref{sec:ftqc}.
We conduct simulations for $\gamma$ values of 0.01 and 1. A $\gamma$ value of 1 signifies the influence of idle qubits on the circuit, while a $\gamma$ value of 0.01 means the idle qubit errors can be mostly ignored. On the other hand, we simulate for $\beta$ values of 1 and 10. Conventionally, the measurement error rate is assumed to be roughly the same as the gate error rate. However, recent experiments suggest that the measurement error rate dominates the gate error rate and can be ten times higher.

We  summarize the number for gate for each type,  number of qubits, circuit depth, and circuit area for one round of flagged SE
for each scheme in  Table~\ref{table.list_area}. 
For   the \codepar{17,1,5} and \codepar{19,1,5} CSS codes,  only the SE circuits of the Z-type stabilizers are accounted for.
For simplicity, we make the following assumptions:
(1) The circuit depth is measured as the longest CNOT or CZ depth.
(2) Ancilla qubits can be reused, so the number of physical qubits for a scheme is the maximum number of qubits required at any instance.
(3) The number of idle qubits is calculated as the circuit depth times the number of qubits minus twice the number of two-qubit gates.

In practice, the circuit area can be optimized; for example, some state preparations can be delayed and some measurements can be done earlier. In the serial $\mathbf{1}^T$ schemes, the ancillary qubits will be measured and reset, but we omit these steps here.

 In the following simulated performance curves of physical gate error rate 
$p$ versus logical error rate $p_L$, we collect at least 1600 logical errors for each physical error rate using Monte Carlo simulations.
	
	Recall that the error threshold considered in Def.~\ref{def:threshold} compares the logical error rate of an encoded codeword with the physical error rate of the dominant CNOT/CZ gates. Therefore, a simulated pseudo-threshold for a scheme is located at the intersection of its performance curve with the $p=p_L$ line.
	
	As shown in Table~\ref{table.list_area}, the three flag-sharing schemes have the smallest effective circuit areas in all four scenarios considered for the three codes, respectively. Consequently, we expect the flag-sharing schemes will outperform the others in terms of error thresholds.

\subsection{The result of the distance-3 code}

\begin{figure}[htbp]
\centering
\includegraphics[width=1\linewidth]{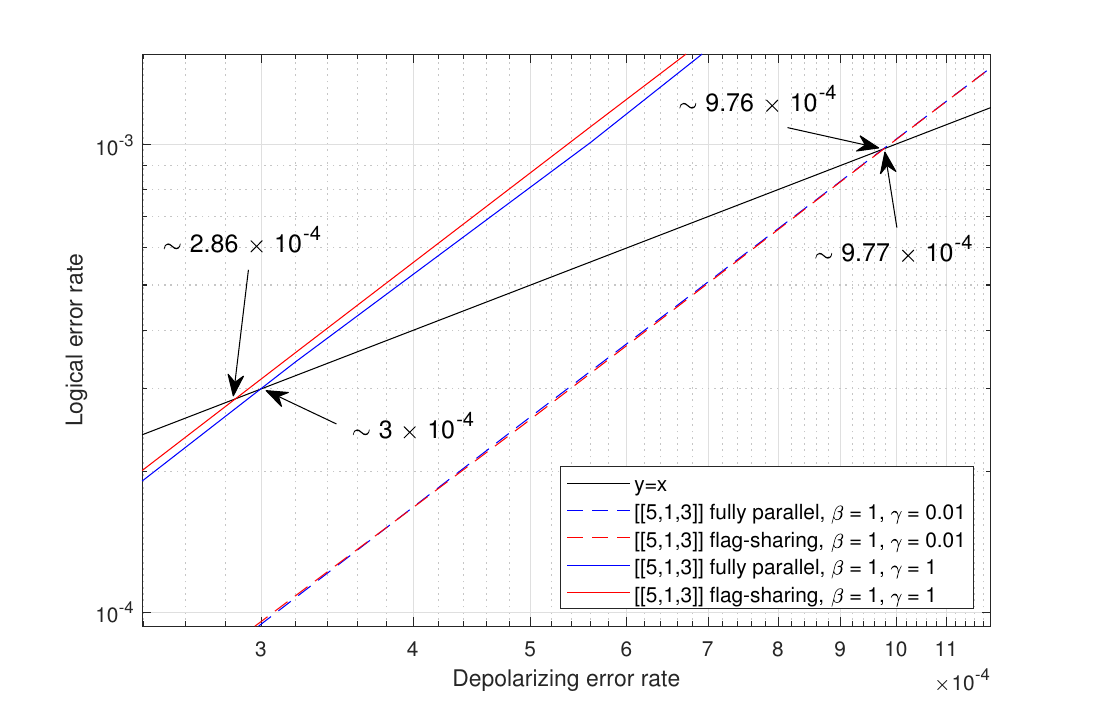}
\caption{ Simulations of the $\begin{bmatrix}
		2\\2
	\end{bmatrix}$  and $\mathbf{1}$ SE schemes for the \codepar{5,1,3} code for $\gamma=0.01$ and $\gamma=1$ at $\beta =1$.}\label{fig.513_result}
\end{figure}

\begin{figure}[htbp]
\centering
\includegraphics[width=1\linewidth]{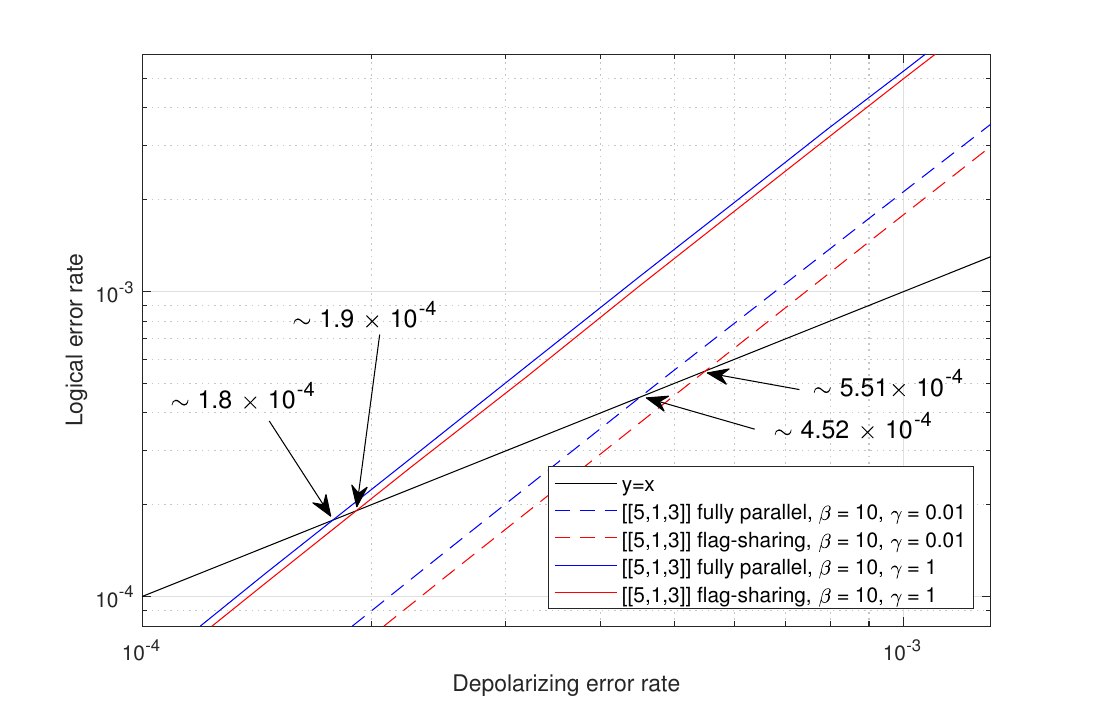}
\caption{  Simulations of the $\begin{bmatrix}
		2\\2
	\end{bmatrix}$  and $\mathbf{1}$ SE schemes for the \codepar{5,1,3} code for $\gamma=0.01$ and $\gamma=1$ at  $\beta =10$.}\label{fig.513_beta_result}
\end{figure}

Employing the decision tree for distance-3 codes~in Fig.~\ref{fig.tree_d3} and the precomputed lookup table~in Table~\ref{tb:513_lookup},
Figures~\ref{fig.513_result} and~\ref{fig.513_beta_result} illustrate the results of simulating different schemes and various values of $\beta$ and $\gamma$ for the \codepar{5,1,3} code.
We have summarized the simulated pseudo-thresholds in Table~\ref{table:d3_result}.

\begin{table}[htbp]
	\begin{center}
		\begin{tabular}[t]{|l|c|c|}
			\hline               
			\multirow{2}{*}{Scheme }         & $\gamma = 0.01$                 & $\gamma = 1$  \\
			\cline{2-3}
											&	 \multicolumn{2}{c|}{$\beta = 1$ }  \\
			\hline
		  serial [1~1~1~1]             &   --       & $7.09 \times 10^{-5}$ \cite{CB18} \\ 
			 flag-sharing $[2~2]^T$     	&  $9.77 \times 10^{-4}$          & $2.86 \times 10^{-4}$  \\
			 fully parallel  $[1~1~1~1]^T$  &  $9.76 \times 10^{-4}$          & $3.0 \times 10^{-4}$  \\
			\hline\hline
			\multirow{2}{*}{Scheme }         & $\gamma = 0.01$                 & $\gamma = 1$  \\
			\cline{2-3}
											&	 \multicolumn{2}{c|}{ {$\beta = 10$} }  \\
			\hline
		  flag-sharing  $[2~2]^T$   	&  $5.51 \times 10^{-4}$          & $1.9 \times 10^{-4}$  \\
			 fully parallel  $[1~1~1~1]^T$  &  $4.52 \times 10^{-4}$          & $1.8 \times 10^{-4}$  \\
			\hline
		\end{tabular}
		\caption{Pseudo-thresholds for   the  flag-sharing $\begin{bmatrix}
				2\\2
			\end{bmatrix}$  and the  fully-parallel $\mathbf{1}$  SE schemes for the \codepar{5,1,3} code.
}\label{table:d3_result}
	\end{center}
\end{table}

The effective circuit areas for the $\begin{bmatrix}
	2\\2
\end{bmatrix}$ flag-sharing scheme and the fully parallel $\mathbf{1}$ scheme are similar, resulting in comparable overall performances.
At $\beta=1$, it can be observed that the  $\begin{bmatrix}
	2\\2
\end{bmatrix}$   scheme performs slightly better than the   $\mathbf{1}$ scheme at $\gamma= 0.01$.
At $\gamma= 1$, although the number of idle qubits in the $\begin{bmatrix}
	2\\2
\end{bmatrix}$ scheme is only slightly higher than in the $\mathbf{1}$ scheme, the CNOT structure in the shared-flag scheme is significantly more complex, leading to a slightly poorer performance.

Both the $\begin{bmatrix}
	2\\2
\end{bmatrix}$  and  $\mathbf{1}$ have better  pseudo-threshold at $\gamma=1$ than the traditional serial flag scheme.

When considering the case $\beta=10$, where the measurement errors are the worst error sources, 
the $\begin{bmatrix}
	2\\2
\end{bmatrix}$ scheme outperforms the $\mathbf{1}$ scheme  for both $\gamma=0.01$ and $\gamma=1$.
This reflects the fact that the $\begin{bmatrix}
	2\\2
\end{bmatrix}$ scheme uses $25\%$ less measurements than the   $\mathbf{1}$ scheme.

\begin{table}[htbp]
	\begin{center}
		\begin{tabular}[t]{|l|c|c|}
			\hline               
			\multirow{2}{*}{Scheme }         & $\gamma = 0.01$                 & $\gamma = 1$  \\
			\cline{2-3}
			&	 \multicolumn{2}{c|}{$\beta = 1$ }  \\
			\hline
			\codepar{7,1,3} parallel \cite{LL23}  &  $1.25 \times 10^{-3}$    &  $1.75 \times 10^{-4}$ \\
			\codepar{9,1,3} parallel \cite{LL23}  &  $8.81 \times 10^{-3}$    &  $8.77 \times 10^{-4}$ \\
			Bacon-Shor-13 \cite{LMB18}           &  $8.09 \times 10^{-3}$    &  $6.41 \times 10^{-4}$ \\
			
			\hline\hline
			\multirow{2}{*}{Scheme }         & $\gamma = 0.01$                & $\gamma = 1$  \\
			\cline{2-3}
			&	 \multicolumn{2}{c|}{ {$\beta = 10$} }  \\
			\hline	
			\codepar{7,1,3} parallel   &  $9.82 \times 10^{-4}$    & $1.6 \times 10^{-4}$  \\
			\codepar{9,1,3} parallel   &  $5.07 \times 10^{-3}$    & $8.24 \times 10^{-4}$  \\
			Bacon-Shor-13           &  $2.71 \times 10^{-3}$    & $4.24 \times 10^{-4}$  \\
			
			\hline
		\end{tabular}
		\caption{{Pseudo-thresholds for various distance-3 code.}} \label{tb:threshold_other_d3}
	\end{center}
\end{table}

For reference, we have also simulated  other distance-3 codes in Table~\ref{tb:threshold_other_d3}.
We can see that at $\gamma=1$,  the \codepar{5,1,3} code emerges as a competitive candidate for near-term experiments, as it has a comparable pseudo-threshold to the \codepar{7,1,3} and \codepar{9,1,3} codes, while using fewer qubits.

\subsection{The result of the distance-5 code}

\begin{figure}[htbp]
	\centering
	\includegraphics[width=1\linewidth]{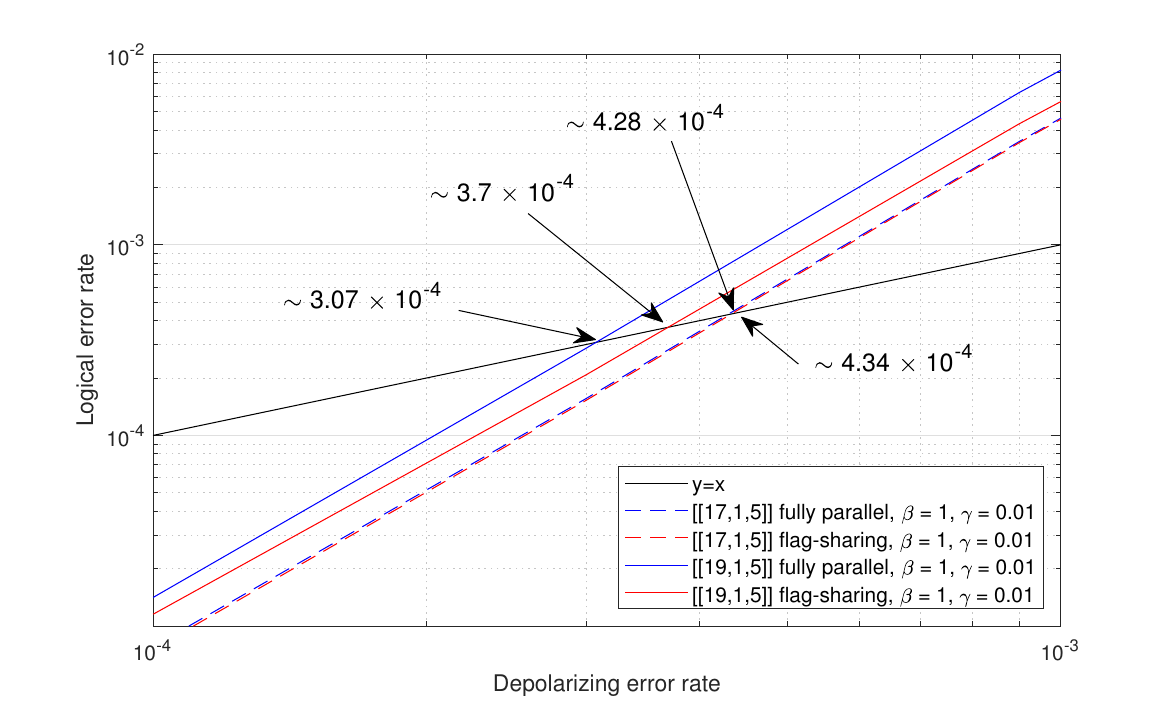}
	\caption{ 
		Simulations of the $\begin{bmatrix}
			2&2&2&1&1
		\end{bmatrix}^T$  and $\mathbf{1}$ SE schemes for the \codepar{17,1,5} code and the $\begin{bmatrix}
		2&2&2&1&1&1
	\end{bmatrix}^T$  and $\mathbf{1}$ SE schemes for the \codepar{19,1,5} code at $\gamma=0.01$ and $\beta =1$.}\label{fig.d5_result1}
\end{figure}
\begin{figure}[htbp]
	\centering
	\includegraphics[width=1\linewidth]{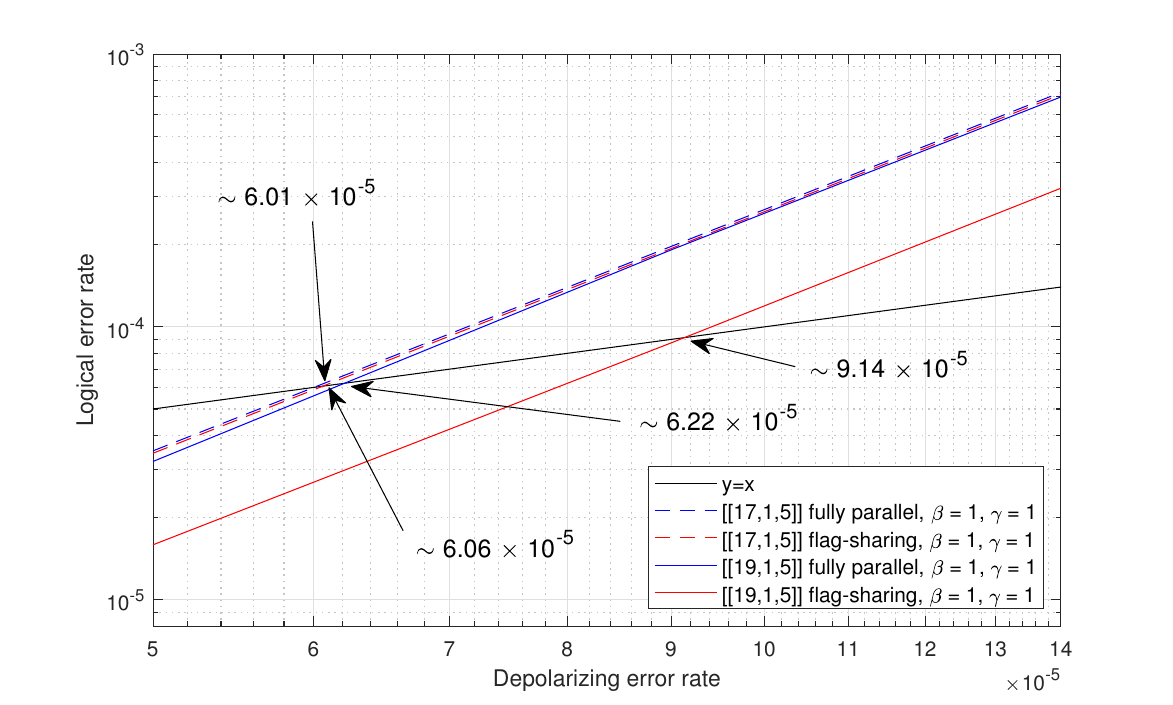}
	\caption{Simulations of the $\begin{bmatrix}
			2&2&2&1&1
		\end{bmatrix}^T$  and $\mathbf{1}$ SE schemes for the \codepar{17,1,5} code and the $\begin{bmatrix}
			2&2&2&1&1&1
		\end{bmatrix}^T$  and $\mathbf{1}$ SE schemes for the \codepar{19,1,5} code at $\gamma=1$ and $\beta =1$.}\label{fig.d5_result2}
\end{figure}

\begin{figure}[htbp]
	\centering
	\includegraphics[width=1\linewidth]{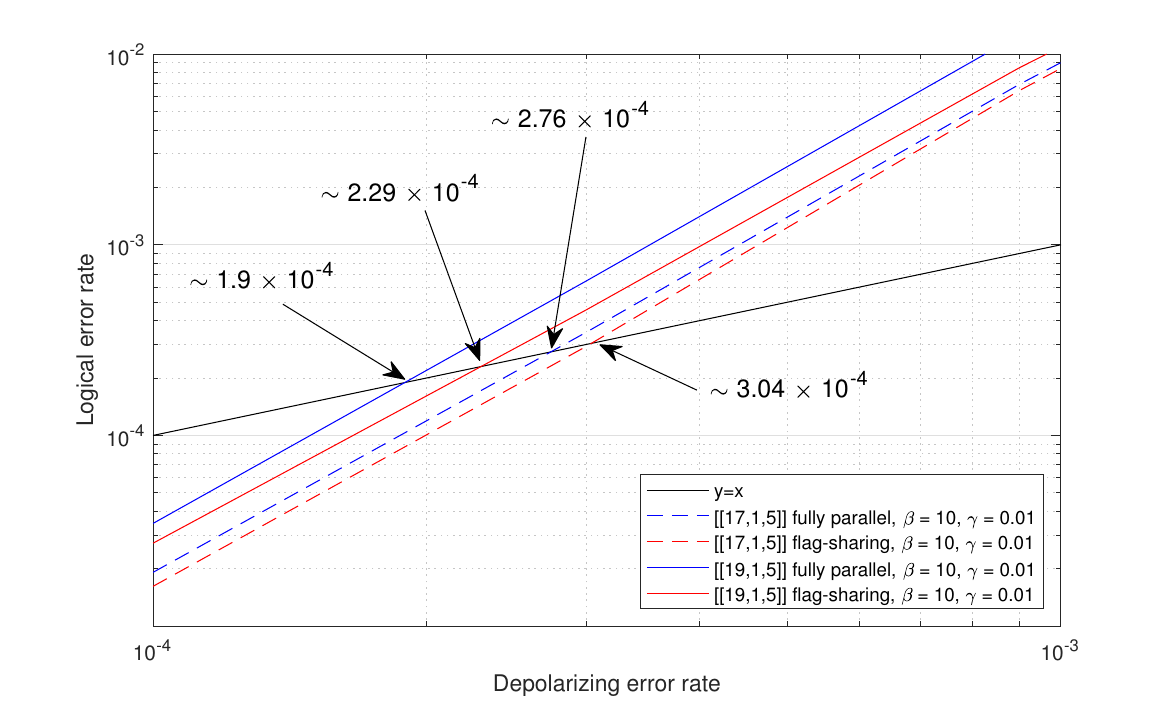}
	\caption{ Simulations of the $\begin{bmatrix}
			2&2&2&1&1
		\end{bmatrix}^T$  and $\mathbf{1}$ SE schemes for the \codepar{17,1,5} code and the $\begin{bmatrix}
			2&2&2&1&1&1
		\end{bmatrix}^T$  and $\mathbf{1}$ SE schemes for the \codepar{19,1,5} code at $\gamma=0.01$ and $\beta =10$.}\label{fig.d5_beta_result1}
\end{figure}

\begin{figure}[htbp]
	\centering
	\includegraphics[width=1\linewidth]{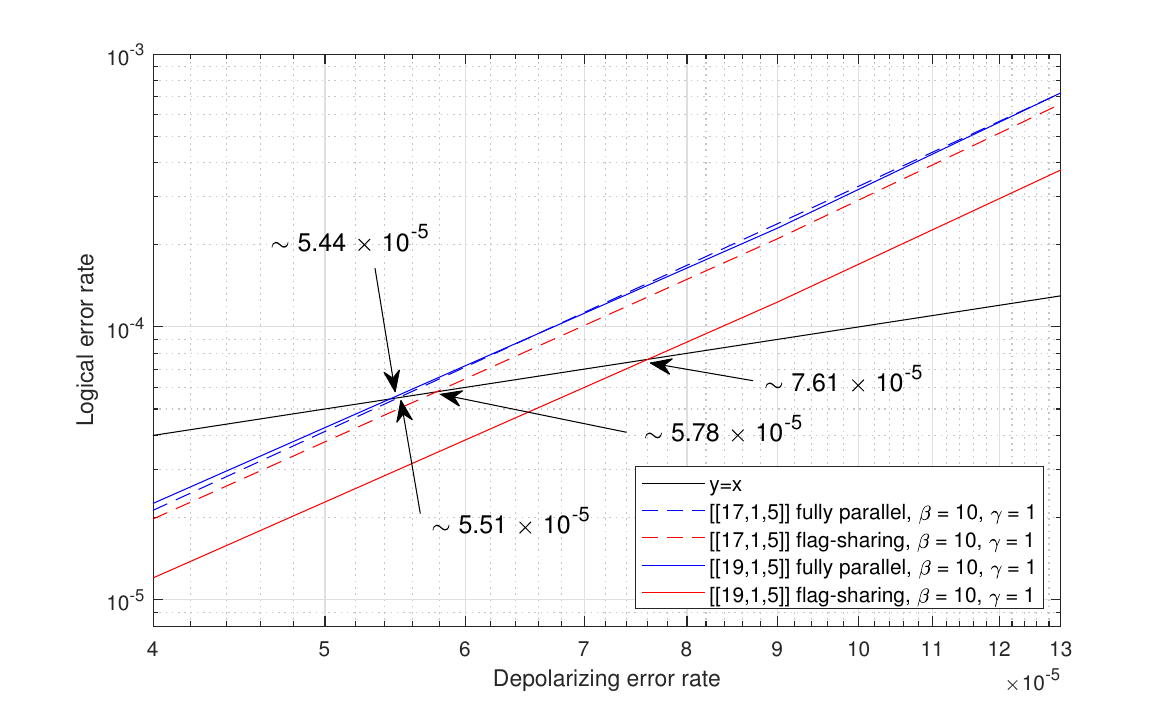}
	\caption{Simulations of the $\begin{bmatrix}
			2&2&2&1&1
		\end{bmatrix}^T$  and $\mathbf{1}$ SE schemes for the \codepar{17,1,5} code and the $\begin{bmatrix}
			2&2&2&1&1&1
		\end{bmatrix}^T$  and $\mathbf{1}$ SE schemes for the \codepar{19,1,5} code at $\gamma=1$ and $\beta =10$.}\label{fig.d5_beta_result2}
\end{figure}

\begin{table}[htbp]
	%\begin{center}
	\begin{tabular}[t]{|l|c|c|}
		\hline               
		\multirow{2}{*}{Scheme }         & $\gamma = 0.01$                 & $\gamma = 1$  \\
		\cline{2-3}
		&	 \multicolumn{2}{c|}{$\beta = 1$ }  \\
		\hline
		\codepar{17,1,5} serial \cite{DML+24}    & --                         &  $5.497 \times 10^{-6}$ \\
		\codepar{17,1,5} [4~4] \cite{DML+24}& --                         &  $7.067 \times 10^{-6}$ \\		
		\codepar{17,1,5} $[2~2~2~1~1]^T$           & $4.34 \times 10^{-4}$      &  $6.06 \times 10^{-5}$ \\
		\codepar{17,1,5} fully parallel  	   & $4.28 \times 10^{-4}$      &  $6.01 \times 10^{-5}$ \\
		\hline
		\codepar{19,1,5} serial \cite{CB18}      &   --                       & $1.14 \times 10^{-5}$  \\ 		
		\codepar{19,1,5} serial \cite{DML+24} 	  &   --                    & $1.183 \times 10^{-5}$ \\ 	
		\codepar{19,1,5} [3~3~3] \cite{DML+24}&   --                       & $1.992 \times 10^{-5}$ \\ 					
		\codepar{19,1,5} $[2~2~2~1~1~1]^T$           & $3.7 \times 10^{-4}$       & $9.14 \times 10^{-5}$ \\
		\codepar{19,1,5} fully parallel        & $3.07 \times 10^{-4}$      & $6.22 \times 10^{-5}$ \\
		\hline\hline
		\multirow{2}{*}{Scheme }         & $\gamma = 0.01$                 & $\gamma = 1$  \\
		\cline{2-3}
		&	 \multicolumn{2}{c|}{$\beta = 10$}  \\
		\hline
		\codepar{17,1,5} $[2~2~2~1~1]^T$            & $ 3.04 \times 10^{-4}$  & $ 5.78 \times 10^{-5}$  \\
		\codepar{17,1,5} fully parallel  		  & $ 2.76 \times 10^{-4}$  & $ 5.51 \times 10^{-5}$  \\	
		\hline
		\codepar{19,1,5} $[2~2~2~1~1~1]^T$           & $ 2.29 \times 10^{-4}$     & $ 7.61 \times 10^{-5}$  \\
		\codepar{19,1,5} fully parallel           & $ 1.9 \times 10^{-4}$      & $ 5.44 \times 10^{-5}$  \\								
		\hline		
	\end{tabular}
	\caption{Pseudo-thresholds for various schemes of the \codepar{17,1,5} and   \codepar{19,1,5} codes.
	}\label{table:d5_result}
	%\end{center}
\end{table}

We utilize the decision tree for distance-5 codes in~Fig.~\ref{fig.tree_d5} to simulate 
the flag-sharing  $\begin{bmatrix}
	2&2&2&1&1
\end{bmatrix}^T$  and fully parallel $\mathbf{1}$ SE schemes for the \codepar{17,1,5} code and the flag-sharing  $\begin{bmatrix}
	2&2&2&1&1&1
\end{bmatrix}^T$  and fully parallel $\mathbf{1}$ SE schemes for the \codepar{19,1,5} code.
The simulation results for various scenarios are shown in Figs.~\ref{fig.d5_result1},~\ref{fig.d5_result2},~\ref{fig.d5_beta_result1}, and \ref{fig.d5_beta_result2}.
We have summarized the simulated pseudo-thresholds in~Table \ref{table:d5_result}.

First, let us discuss the SE schemes for the \codepar{17,1,5} codes at $\gamma=1$ and $\beta=1$. From Table~\ref{table.list_area}, both our flag-sharing and fully parallel schemes exhibit pseudo-thresholds almost an order of magnitude better than the serial scheme and the [4~4] scheme in~\cite{DML+24}. This is expected given the effective circuit areas  	 in Table~\ref{table.list_area}, where both our flag-sharing and fully parallel schemes use less than half the area of the serial scheme in a single round of SE. Since three or four rounds of SE are performed in general, the effects are further amplified, resulting in much better pseudo-thresholds for our schemes.

Similarly, both our flag-sharing and fully parallel schemes for the \codepar{19,1,5} codes demonstrate pseudo-thresholds three to eight times better than the serial scheme and the [3~3~3] scheme in~\cite{DML+24}. These results highlight the effectiveness of our method, using flag-sharing and parallelism with generalized error syndromes to reduce the effective circuit areas.

At $\beta=1$, the 
the flag-sharing  $\begin{bmatrix}
	2&2&2&1&1
\end{bmatrix}^T$  scheme performs slightly better than the fully parallel $\mathbf{1}$ scheme for the \codepar{17,1,5} code 
since its effective circuit area is smaller.
At $\beta=10$,  the gap between the two schemes is amplified since the flag-sharing scheme uses $1/6$ less measurements.

In summary, both our flag-sharing schemes perform better than the  fully parallel schemes for the \codepar{17,1,5} and \codepar{19,1,5} codes,
making them the best candidates for FT SE among the scenarios we have considered.

\section{Conclusion}\label{sec.conclusion}

We proposed both sequential flag-sharing and fully parallel SE schemes for general CSS stabilizer codes, outperforming traditional serial flag schemes and other structural types. Various flag-sharing schemes can be realized through different forms of tiling, providing a flexible method in the design of fault-tolerant SE circuits. In addition,  we introduced the notion of effective circuit area, serving as a rough but quick tool for evaluating circuit quality.

By constructing a decision tree for decoding, we obtained a generalized error syndrome table with an enlarged syndrome space. However, most of these generalized syndrome spaces are not used, leaving room for optimization. For instance, we did not construct a table of $2^{10}$ corrections for the \codepar{5,1,3} code, suggesting potential optimization to handle some 2-fault events. Similar opportunities for optimization exist for the \codepar{17,1,5} and \codepar{19,1,5} codes.

We did not find a $[2~2~2~2]$ SE scheme for the \codepar{17,1,5} code, which, if it exists, may imply an improved threshold over the flag-sharing $[2~2~2~1~1]^T$ scheme.
Similar discussion applies to the \codepar{19,1,5} code as well.

We also presented versatile SE schemes for the \codepar{5,1,3} code. By using a different set of stabilizer generators and generalized error syndromes, we created an SE circuit featuring the smallest circuit area reported in the literature. Moreover, we utilized the cyclic structure of the stabilizer generators to construct versatile SE schemes for the \codepar{5,1,3} code, which can be generalized to other non-CSS codes.

Using generalized error syndromes for lookup tables, we constructed decision trees simplified from those in \cite{CB18}. The decisions we presented in Figs.~\ref{fig.tree_d3} and \ref{fig.tree_d5} can be recursively extended for codes with higher minimum distances. By using all the measurement outcomes as the index for the lookup table, we presented the decision tree in a compressed manner. It is worth exploring the rapid and automated construction of syndrome tables or developing alternative types of decoders.

In \cite{BSE+23}, an idea was proposed for improving a fault-tolerant procedure based on a lookup table-based decoder, suggesting that some stabilizers do not need to be measured if it can be ensured they commute with the errors when certain flag qubits are raised. This idea could further improve the simulated pseudo-thresholds for our schemes.

The source files of our computer programs can be found in Ref. \cite{LL24code}.
 
\bibliography{qecc}

\appendix

\section{Syndrome extraction schemes for the \codepar{17,1,5} code}\label{app:1715}

\begin{figure*}[htbp]
	\centering
	\includegraphics[width=1\linewidth]{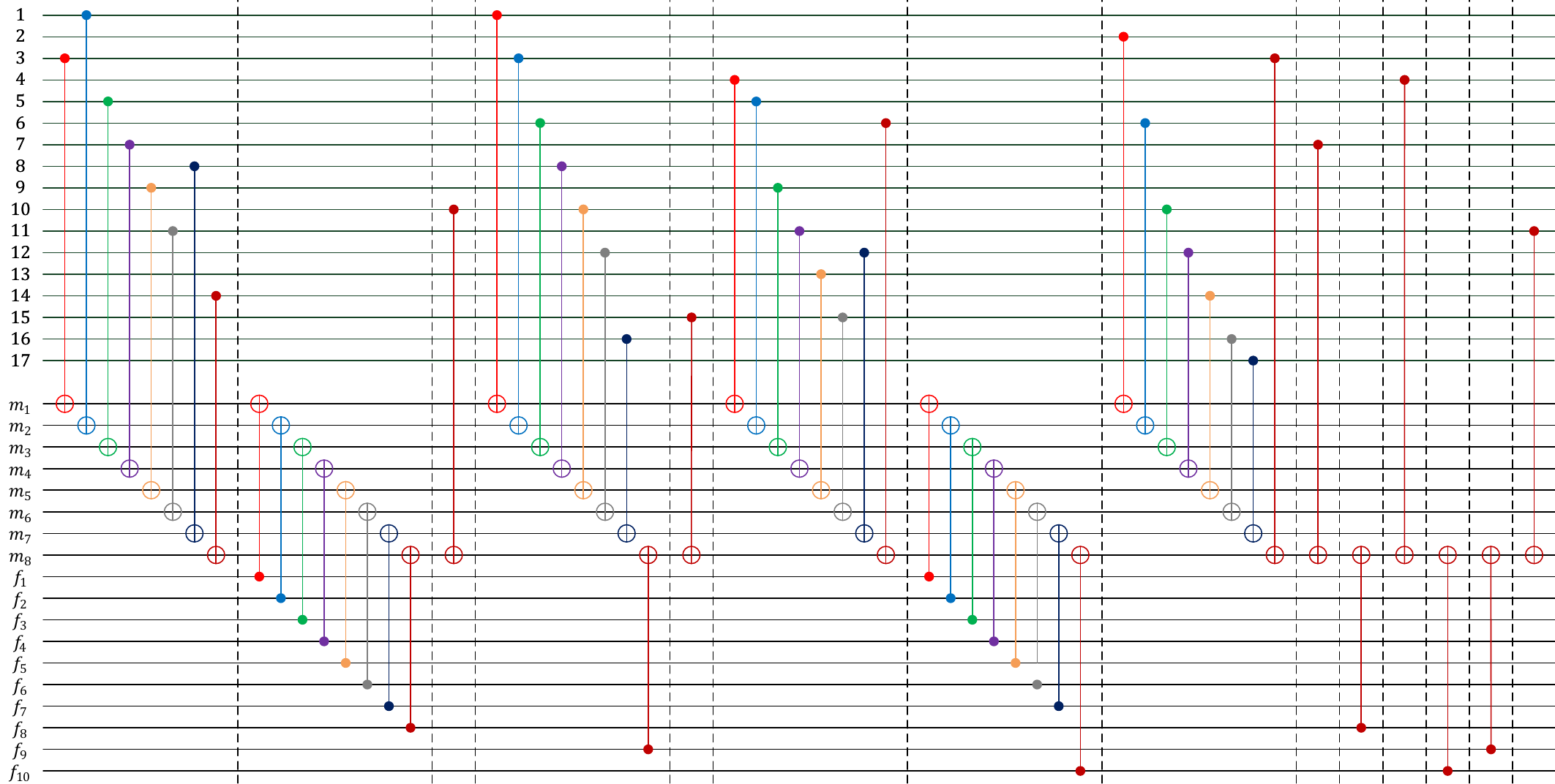}
	\caption{A fully parallel $\mathbf{1}$ SE circuit for the $Z$-type stabilizers of the \codepar{17,1,5} code.
		Each stabilizer measurement is represented by a distinct color.
		Each stabilizer measurement has an independent set of ancilla and flag qubits. 
		Note that, for space-saving purpose, we have omitted state preparations and measurements.
	}\label{fig.1715_full}
\end{figure*}

\begin{figure*}[htbp]
	\centering
	\includegraphics[width=1\linewidth]{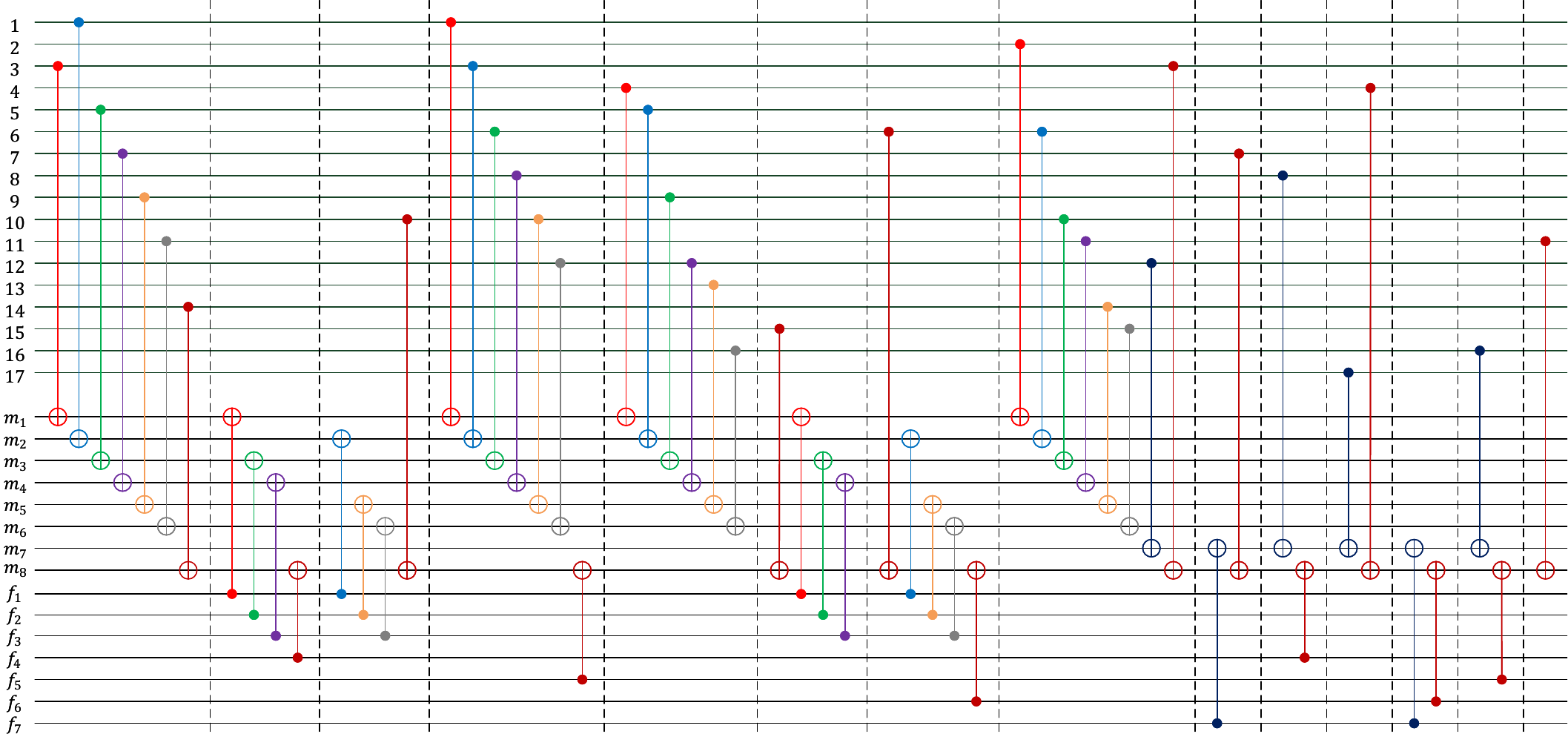}
	\caption{ A parallel $[2;~2;~2;~1;~1]$ flag-sharing SE circuit for the $Z$-type stabilizers of the $\codepar{17,1,5}$ code.
		Note that $g_1$ and $g_2$ share a flag qubit $f_1$,  $g_3$ and $g_5$ share a flag qubit $f_2$, $g_4$ and the $g_6$ share a flag qubit $f_3$. The stabilizer $g_7$ uses the flag qubit $f_7$,
		while the stabilizer $g_8$ uses the flag qubits $f_4$, $f_5$, and  $f_6$. 
	}\label{fig.1715_parallel}
\end{figure*}

We provide the detailed circuits for the  $\mathbf{1}$  SE scheme and the $[2;~2;~2;~1;~1]$ flag-sharing SE scheme
for the \codepar{17,1,5} code in Figs.~\ref{fig.1715_full} and~\ref{fig.1715_parallel}, respectively.

\section{Partial Lookup table for the \codepar{5,1,3} code}\label{sec.appendix.513}

In this section, we provide a partial lookup table in Table~\ref{tb:513_lookup} for the $\begin{bmatrix}
	2;2
\end{bmatrix}$ scheme for the \codepar{5,1,3} code.

\begin{table*}[htbp]
	\begin{scriptsize}
		\begin{tabular}{cc}
			\begin{tabular}{|c|c|c|c|c|}
				\hline
				
				Failure  & Data $X$ error & Data $Z$ error & $m_{1534}f_{12} $ & $m^{\prime}_{1534}$    \\
				\hline

				$A_{IX}$ & 00101 & 00000 & 000101  & 1011        \\	
				$A_{XX}$ & 01101 & 00000 & 000101  & 0111        \\	
				$A_{ZX}$ & 00101 & 01000 & 000001  & 1010        \\	
				$A_{YX}$ & 01101 & 01000 & 000001  & 0110        \\
				
				$B_{IX}$ & 10010 & 00000 & 001010  & 0111        \\		
				$B_{XX}$ & 10110 & 00000 & 001010  & 1111        \\	
				$B_{ZX}$ & 10010 & 00100 & 011010  & 0001        \\	
				$B_{YX}$ & 10110 & 00100 & 011010  & 1001        \\	
				
				$C_{IX}$ & 01000 & 00001 & 000001  & 1000        \\		
				$C_{XX}$ & 11000 & 00001 & 000001  & 1101        \\	
				$C_{ZX}$ & 01000 & 10001 & 101001  & 0010        \\	
				$C_{YX}$ & 11000 & 10001 & 101001  & 0111        \\
				
				$D_{IX}$ & 00001 & 00000 & 000101  & 0011        \\	
				$D_{XX}$ & 00101 & 00000 & 000101  & 1011        \\	
				$D_{ZX}$ & 00001 & 00100 & 000101  & 0101        \\	
				$D_{YX}$ & 00101 & 00100 & 000101  & 1101        \\		
				
				$E_{IX}$ & 10000 & 00000 & 000010  & 0101        \\		
				$E_{XX}$ & 10010 & 00000 & 001010  & 0111        \\	
				$E_{ZX}$ & 10000 & 00010 & 000010  & 1100        \\	
				$E_{YX}$ & 10010 & 00010 & 001010  & 1110        \\

				$F_{IX}$ & 10000 & 00010 & 000010  & 1100        \\	
				$F_{XX}$ & 10001 & 00010 & 000110  & 1111        \\	
				$F_{ZX}$ & 10000 & 00011 & 010010  & 1000        \\	
				$F_{YX}$ & 10001 & 00011 & 010110  & 1011        \\	
				
				$G_{IX}$ & 00000 & 00010 & 000010  & 1001        \\			
				$G_{XX}$ & 10000 & 00010 & 000010  & 1100        \\	
				$G_{ZX}$ & 00000 & 10010 & 100010  & 0011        \\	
				$G_{YX}$ & 10000 & 10010 & 100010  & 0110        \\		
				
				$H_{IX}$ & 00000 & 00001 & 000001  & 0100        \\		
				$H_{XX}$ & 01000 & 00001 & 000001  & 1000        \\	
				$H_{ZX}$ & 00000 & 01001 & 000001  & 0101        \\	
				$H_{YX}$ & 01000 & 01001 & 000001  & 1001        \\	
				
				$a_{XI}$ & 10010 & 00100 & 011010  & 0001        \\			
				$a_{XX}$ & 10010 & 00100 & 111010  & 0001        \\	
				$a_{XZ}$ & 10010 & 00100 & 011000  & 0001        \\	
				$a_{XY}$ & 10010 & 00100 & 111000  & 0001        \\	
				$b_{XI}$ & 10000 & 00011 & 010010  & 1000        \\			
				$b_{XX}$ & 10000 & 00011 & 111010  & 1000        \\	
				$b_{XZ}$ & 10000 & 00011 & 010000  & 1000        \\	
				$b_{XY}$ & 10000 & 00011 & 111000  & 1000        \\	
				$c_{XI}$ & 10000 & 00000 & 000000  & 0101        \\		
				$c_{XX}$ & 10000 & 00000 & 001000  & 0101        \\	
				$c_{XZ}$ & 10000 & 00000 & 000010  & 0101        \\	
				$c_{XY}$ & 10000 & 00000 & 001010  & 0101        \\	
				$d_{XI}$ & 00000 & 00010 & 000000  & 1001        \\								
				$d_{XX}$ & 00000 & 00010 & 000000  & 1001        \\	
				$d_{XZ}$ & 00000 & 00010 & 000010  & 1001        \\	
				$d_{XY}$ & 00000 & 00010 & 000010  & 1001        \\

				$e_{XI}$ & 00101 & 01000 & 000001  & 1010        \\			
				$e_{XX}$ & 00101 & 01000 & 010001  & 1010        \\	
				$e_{XZ}$ & 00101 & 01000 & 000000  & 1010        \\	
				$e_{XY}$ & 00101 & 01000 & 010000  & 1010        \\		
				$f_{XI}$ & 01000 & 10001 & 101001  & 0010        \\		
				$f_{XX}$ & 01000 & 10001 & 111101  & 0010        \\	
				$f_{XZ}$ & 01000 & 10001 & 101000  & 0010        \\	
				$f_{XY}$ & 01000 & 10001 & 111100  & 0010        \\	
				$g_{XI}$ & 00001 & 00000 & 000100  & 0011        \\		
				$g_{XX}$ & 00001 & 00000 & 000000  & 0011        \\	
				$g_{XZ}$ & 00001 & 00000 & 000101  & 0011        \\	
				$g_{XY}$ & 00001 & 00000 & 000001  & 0011        \\	
				$h_{XI}$ & 00000 & 00001 & 000000  & 0100        \\								
				$h_{XX}$ & 00000 & 00001 & 000000  & 0100        \\	
				$h_{XZ}$ & 00000 & 00001 & 000001  & 0100        \\	
				$h_{XY}$ & 00000 & 00001 & 000001  & 0100        \\

				\hline
			\end{tabular}

			&
			
			\begin{tabular}{|c|c|c|c|c|}
				\hline
				
				Failure  & Data $X$ error & Data $Z$ error & $m_{1534}f_{12} $ & $m^{\prime}_{1534}$    \\
				\hline
				$A_{IY}$  & 00101 & 00000 & 010101 & 1011		\\
				$A_{XY}$  & 01101 & 00000 & 010101 & 0111		\\
				$A_{ZY}$  & 00101 & 01000 & 010001 & 1010		\\
				$A_{YY}$  & 01101 & 01000 & 010001 & 0110		\\

				$B_{IY}$ & 10010 & 00000 & 101010 & 0111		\\
				$B_{XY}$ & 10110 & 00000 & 101010 & 1111		\\
				$B_{ZY}$ & 10010 & 00100 & 111010 & 0001		\\
				$B_{YY}$ & 10110 & 00100 & 111010 & 1001		\\
				
				$C_{IY}$  & 01000 & 00001 & 000101 & 1000		\\
				$C_{XY}$  & 11000 & 00001 & 000101 & 1101		\\
				$C_{ZY}$  & 01000 & 10001 & 101101 & 0010		\\
				$C_{YY}$  & 11000 & 10001 & 101101 & 0111		\\
				
				$D_{IY}$  & 00001 & 00000 & 010101 & 0011		\\
				$D_{XY}$  & 00101 & 00000 & 010101 & 1011		\\
				$D_{ZY}$  & 00001 & 00100 & 010101 & 0101		\\
				$D_{YY}$  & 00101 & 00100 & 010101 & 1101		\\
				
				$E_{IY}$ & 10000 & 00000 & 100010 & 0101		\\
				$E_{XY}$ & 10010 & 00000 & 101010 & 0111		\\
				$E_{ZY}$ & 10000 & 00010 & 100010 & 1100		\\
				$E_{YY}$ & 10010 & 00010 & 101010 & 1110		\\
				
				$F_{IY}$ & 10000 & 00010 & 001010 & 1100		\\
				$F_{XY}$ & 10001 & 00010 & 001110 & 1111		\\
				$F_{ZY}$ & 10000 & 00011 & 011010 & 1000		\\
				$F_{YY}$ & 10001 & 00011 & 011110 & 1011		\\
				
				$G_{IY}$ & 00000 & 00010 & 001010 & 1001		\\
				$G_{XY}$ & 10000 & 00010 & 001010 & 1100		\\
				$G_{ZY}$ & 00000 & 10010 & 101010 & 0011		\\
				$G_{YY}$ & 10000 & 10010 & 101010 & 0110		\\

				$H_{IY}$  & 00000 & 00001 & 000101 & 0100		\\
				$H_{XY}$  & 01000 & 00001 & 000101 & 1000		\\
				$H_{ZY}$  & 00000 & 01001 & 000101 & 0101		\\
				$H_{YY}$  & 01000 & 01001 & 000101 & 1001		\\
				
				$a_{YI}$ & 10010 & 00100 & 111010 & 0001		\\
				$a_{YX}$ & 10010 & 00100 & 011010 & 0001		\\
				$a_{YZ}$ & 10010 & 00100 & 111000 & 0001		\\
				$a_{YY}$ & 10010 & 00100 & 011000 & 0001		\\
				$b_{YI}$ & 10000 & 00011 & 011010 & 1000		\\
				$b_{YX}$ & 10000 & 00011 & 110010 & 1000		\\
				$b_{YZ}$ & 10000 & 00011 & 011000 & 1000		\\
				$b_{YY}$ & 10000 & 00011 & 110000 & 1000		\\
				$c_{YI}$ & 10000 & 00000 & 100000 & 0101		\\
				$c_{YX}$ & 10000 & 00000 & 101000 & 0101		\\
				$c_{YZ}$ & 10000 & 00000 & 100010 & 0101		\\
				$c_{YY}$ & 10000 & 00000 & 101010 & 0101		\\
				$d_{YI}$ & 00000 & 00010 & 001000 & 1001		\\
				$d_{YX}$ & 00000 & 00010 & 001000 & 1001		\\
				$d_{YZ}$ & 00000 & 00010 & 001010 & 1001		\\
				$d_{YY}$ & 00000 & 00010 & 001010 & 1001		\\

				$e_{YI}$  & 00101 & 01000 & 010001 & 1010		\\
				$e_{YX}$  & 00101 & 01000 & 000001 & 1010		\\
				$e_{YZ}$  & 00101 & 01000 & 010000 & 1010		\\
				$e_{YY}$  & 00101 & 01000 & 000000 & 1010		\\
				$f_{YI}$  & 01000 & 10001 & 101101 & 0010		\\
				$f_{YX}$  & 01000 & 10001 & 111001 & 0010		\\
				$f_{YZ}$  & 01000 & 10001 & 101100 & 0010		\\
				$f_{YY}$  & 01000 & 10001 & 111000 & 0010		\\
				$g_{YI}$  & 00001 & 00000 & 010100 & 0011		\\
				$g_{YX}$  & 00001 & 00000 & 010000 & 0011		\\
				$g_{YZ}$  & 00001 & 00000 & 010101 & 0011		\\
				$g_{YY}$  & 00001 & 00000 & 010001 & 0011		\\
				$h_{YI}$  & 00000 & 00001 & 000100 & 0100		\\
				$h_{YX}$  & 00000 & 00001 & 000100 & 0100		\\
				$h_{YZ}$  & 00000 & 00001 & 000101 & 0100		\\
				$h_{YY}$  & 00000 & 00001 & 000101 & 0100		\\	
				
				\hline
			\end{tabular}
			
		\end{tabular}
	\end{scriptsize}
	\caption{The lookup table for the  $\begin{bmatrix}
			2\\2
		\end{bmatrix}$ SE scheme for the \codepar{5,1,3} code of Fig.~\ref{fig.513_parallel_parallel_v3}. The entry $m_{1534} f_{12}$ denotes the measurement outcomes of the first round of the $\begin{bmatrix}
		2\\2
	\end{bmatrix}$ SE, while the entry $m'_{1534}$ denotes the measurement outcomes of the subsequent round of a raw SE.}
\label{tb:513_lookup}
\end{table*}

\section{Syndrome extraction schemes for the \codepar{19,1,5} code}\label{app:1915}

\begin{figure*}[htbp]
	\centering
	\includegraphics[width=1\linewidth]{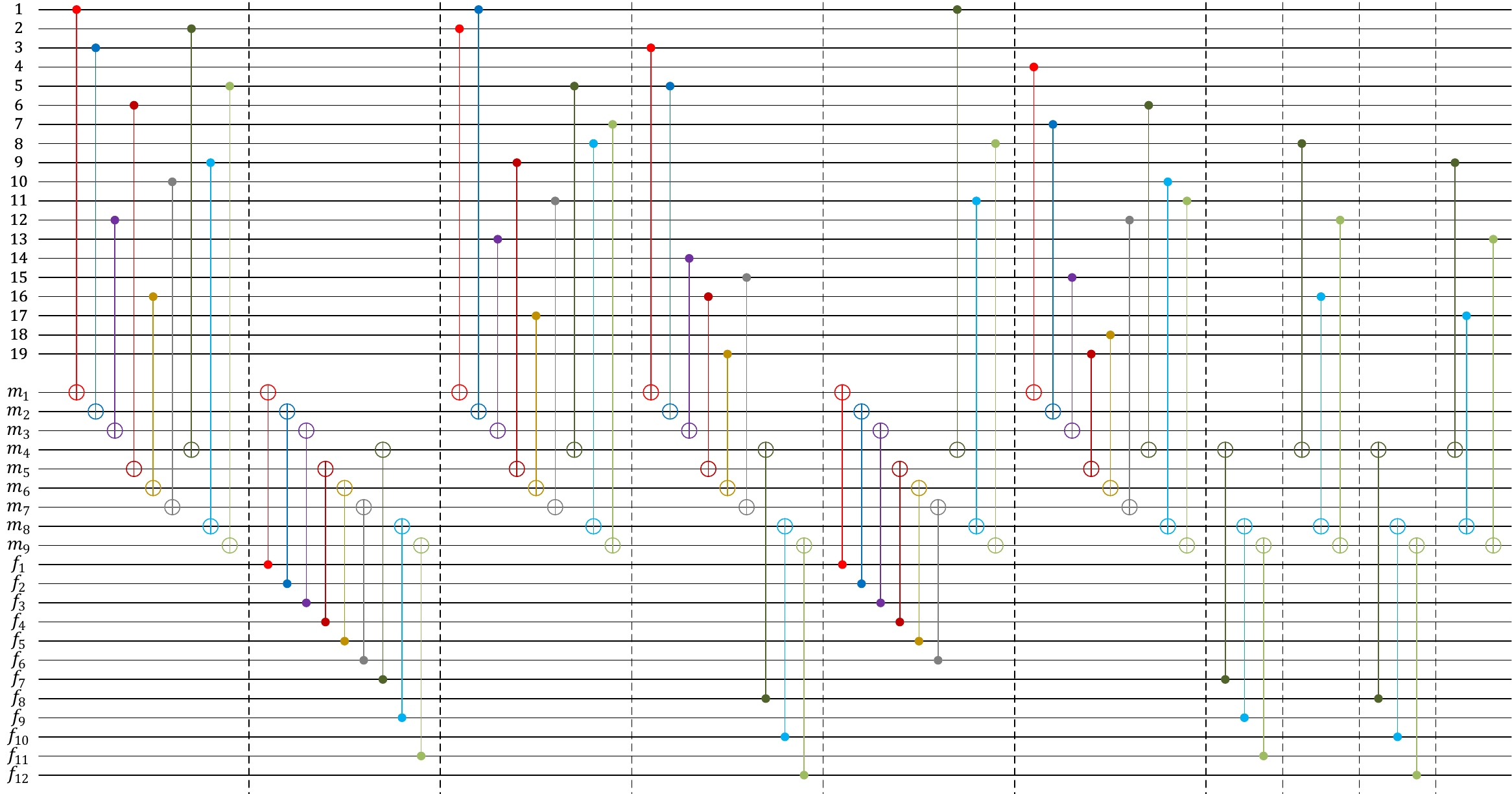}
	\caption{ A $\mathbf{1}$ SE scheme for the $Z$-type stabilizers  of the \codepar{19,1,5} code.
	}\label{fig.1915_full}
\end{figure*}

\begin{figure*}[htbp]
	\centering
	\includegraphics[width=1\linewidth]{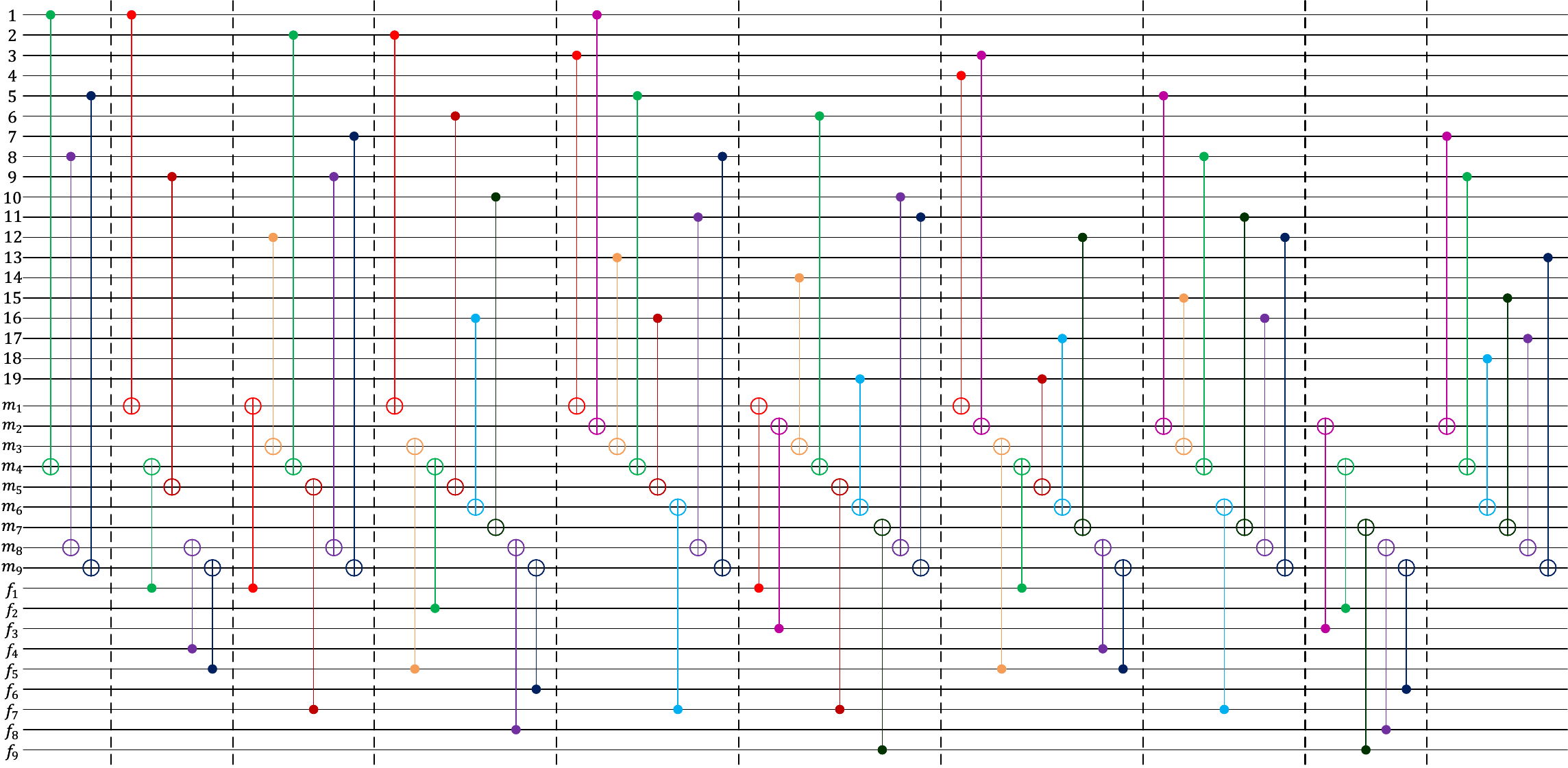}
	\caption{ A parallel $[2;~2;~2;~1;~1;~1]$ SE scheme for the  $Z$-type stabilizers of the \codepar{19,1,5} code.
	}\label{fig.1915_parallel}
\end{figure*}

   We provide the detailed circuits for the  $\mathbf{1}$   SE scheme and the $[2;~2;~2;~1;~1;~1]$ flag-sharing SE scheme
   for the \codepar{19,1,5} code in Figs.~\ref{fig.1915_full} and~\ref{fig.1915_parallel}, respectively.

\end{document}